\documentclass[11pt,draftclsnofoot,journal,onecolumn]{IEEEtran}

\usepackage[dvips]{graphicx}
\usepackage[tight,scriptsize,sf,SF]{subfigure}
\usepackage{amsmath}
\usepackage{amssymb}
\usepackage{url}
\usepackage[nocompress]{cite}

\renewcommand{\IEEEQED}{\IEEEQEDopen}

\begin{document}

\title{Impact of Connection Admission Process on the Direct Retry Load Balancing Algorithm in Cellular Networks}
\author{Przemys{\l}aw Pawe{\l}czak, Shaunak Joshi, Sateesh Addepalli, John Villasenor, and Danijela \v{C}abri{\'c}%
\thanks{Przemys{\l}aw Pawe{\l}czak was with the Department of Electrical Engineering, University of California, Los Angeles. He is currently with Fraunhofer Institute for Telecommunications, Heinrich Hertz Institute, Einsteinufer 37, 10587 Berlin, Germany (email: przemyslaw.pawelczak@hhi.fraunhofer.de).}
\thanks{Shaunak Joshi was with the Department of Electrical Engineering, University of California, Los Angeles. He is currently with Cisco Systems, Inc., San Jose, CA 95134, USA (email: shaunjos@cisco.com).}\thanks{Sateesh Addepalli is with Cisco Systems, Inc., San Jose, CA 95134, USA (email: sateeshk@cisco.com).}
\thanks{John Villasenor and Danijela \v{C}abri{\'c} are with the Department of Electrical Engineering, University of California, Los Angeles, 56-125B Engineering IV Building, Los Angeles, CA 90095-1594, USA (email: \{villa, danijela\}@ee.ucla.edu).}%
\thanks{Part of this work has been accepted to the proceedings of IEEE GLOBECOM, 6-10 Dec., 2010, Miami, FL, USA~\cite{joshi_submitted_2010}.}%
\thanks{Copyright\copyright~2012 IEEE. Personal use of this material is permitted. However, permission to use this material for any other purposes must be obtained from the IEEE by sending a request to pubs-permissions@ieee.org.}}

\maketitle

\begin{abstract}
We present an analytical framework for modeling a priority-based load balancing scheme in cellular networks based on a new algorithm called direct retry with truncated offloading channel resource pool (DR$_{K}$). The model, developed for a baseline case of two cell network, differs in many respects from previous works on load balancing. Foremost, it incorporates the call admission process, through random access. In specific, the proposed model implements the Physical Random Access Channel used in 3GPP network standards. Furthermore, the proposed model allows the differentiation of users based on their priorities. The quantitative results illustrate that, for example, cellular network operators can control the manner in which traffic is offloaded between neighboring cells by simply adjusting the length of the random access phase. Our analysis also allows for the quantitative determination of the blocking probability individual users will experience given a specific length of random access phase. Furthermore, we observe that the improvement in blocking probability per shared channel for load balanced users using DR$_{K}$ is maximized at an intermediate number of shared channels, as opposed to the maximum number of these shared resources. This occurs because a balance is achieved between the number of users requesting connections and those that are already admitted to the network. We also present an extension of our analytical model to a multi-cell network (by means of an approximation) and an application of the proposed load balancing scheme in the context of opportunistic spectrum access.
\end{abstract}

Given the rapid current and expected growth in 3G/UMTS and LTE-based networks and in the number of mobile devices that use such networks to download data-intensive, multimedia-rich content, the need for QoS-enabled connection management is vital. However, the non-uniform distribution of users and consequent imbalance in usage of radio resources leads to an existence of local areas of under- and over-utilization of these resources in the network. This phenomenon results in challenging network management issues. Load balancing is an important technique that attempts to solve such issues, and occurs when a centralized network controller intelligently distributes connections from highly congested cells to neighboring cells which are less occupied. This allows for an increase in network subscriber satisfaction because more subscribers meet their QoS requirements. Furthermore, it allows for an increase in overall channel utilization by leveraging the fact that users with access to multiple cells also have access to additional resources. 

In this work, we aim to quantify the impact of load balancing on the overall system as well as user experience (from the separate viewpoints of users that share resources and those that use these shared resources) using a detailed analytical model of a fundamental two cell setup, later extended by means of approximations to multi-cell setup. Furthermore, we present an application of our model in the context of a cellular system using opportunistic spectrum access, which even further improves the teletraffic properties of the considered cellular network, to efficiently utilize system resources.

\subsection{Related Work}
\label{sec:related_work}

Load balancing has been explored for many years as described, for example, in earlier works such as~\cite{eklundh_ieeetc_1986,everitt_jsac_1989}. More recent works such as ~\cite{tonguz_ieeetmc_2008,wu_ieeemwc_2004, song_ieeenet_2007,yanmaz_jsac_2004} have also examined the impacts of various load balancing schemes. 

For simplicity, the previous studies of load balancing have assumed that the connection admission process can be neglected because non-finite user populations are considered with a particular arrival rate of the number connections. In cellular networks, however, finite user population exists where each new connection needs to first send a request to the serving base station (BS) through some predefined control channel. In 3GPP standards, this control channel is the Physical Random Access Channel~\cite[Sec. 2.4.4.4]{laiho_book},~\cite[Sec. II]{cooper_tvt_2002}. The success of a connection request by a user is dependent on multiple factors including the number of requesting users; the pairwise channel quality between the user and the serving base station (measured, for example, in BER or outage probability); and the actual control channel access technique itself~\cite{yang_ieeetwc_2008,lin_vtc_2000,he_electrlett_2007}. 

More recents works, such as~\cite{son_ieeetwc_2009}, use inter- and intra-cell handover techniques to alleviate issues of coverage for cell-edge users without specific focus on traffic parameters. Other recent studies propose the use of low-power cellular relays in the presence of high-power BSs with more emphasis on a heuristic to determine load balancing and mobile association~\cite{Yu_ieeemcom_2011}. It is evident that current literature on load balancing lacks an analytical model that allows for substantial flexibility in terms of finite user population consideration and the use of various traffic parameters.

\subsection{Our Contribution}
\label{sec:contribution}

Until now the exact impact of random access overhead on load balancing performance is not well understood. More specifically, the quantitative relationship between the random access phase length, the user blocking probability and system channel utilization in a load balancing-enabled cellular system is unknown. To the best of our knowledge an analytical model to quantify system-wide and user experience metrics in this respect has not been previously provided. Our work:
\begin{enumerate}
\item
Through development of an analytical model for a baseline (fundamental) two-cell network, demonstrates the benefits of load balancing, from a teletraffic point of view, using realistic traffic scenarios, various network configurations and parameters, and simplified physical layer model for channel quality;
\item
Provides a detailed connection admission process to determine the effects of a finite user population on the efficiency of load balancing performance metrics; and
\item
Allows extension to more complex network setups. More specifically, we present an approximation to a multi-cell case and also provide an application of the model in the context of cellular opportunistic spectrum access.
\end{enumerate}
Our model can be used in:
\begin{enumerate}
\item
Demonstrating that the random access phase length is an important tool that can be used to control both system-wide and user experience performance metrics, for example, quantifying the tradeoffs between random access channel collision probability and blocking probability for load-balanced users as a function of random access phase length.
\item
Determining the impact of channel quality to improve the accuracy of reported load balancing efficiency; for example, quantifying the loss in accuracy due to perfect channel condition assumptions; and
\item
Exploring the effects of varying shared channel access on system-wide performance and user experience, for example, determining the marginal gain of adding more channels for shared access between cells.
\end{enumerate}

The rest of the paper is organized as follows. The system model is introduced in Section~\ref{sec:system_model}, while the analytical model is introduced in Section~\ref{sec:numerical_analysis}. The numerical results are presented in Section~\ref{sec:numerical_results}. Lastly, the paper is concluded in Section~\ref{sec:conclusions}.

\section{System Model}
\label{sec:system_model}

In the following sections we will describe the system model in detail. We start with a description of the channel structure in Section~\ref{sec:channel_structure}, followed by the description of node placement in Section~\ref{sec:node_placement}. Section~\ref{sec:signal_transmission_model} describes the signal transmission model and Section~\ref{sec:prioritization} prioritization policies in load balancing. Then, in Section~\ref{sec:random_access}, introduces a random access process in the context of cellular networks, followed by the introduction of a data transfer model in Section~\ref{sec:data_transfer}. Finally, the whole load balancing process is introduced in Section~\ref{sec:load_balancing_Process}.

\subsection{Channel Structure}
\label{sec:channel_structure}

We consider a cellular system where two BSs are positioned such that they create a region of overlap in coverage. Naturally, while cellular systems typically have far more than two BSs, a reduction to a two-BS system for analytical reasons enables a tractable analytical framework while still allowing exploration of a large number of microscopic parameters to use in optimizing network performance. Furthermore, two-cells provide a fundamental functional pair for the purposes of studying load balancing from the context of off-loading connections from a highly congested cell to a less congested one. We strongly emphasize that consideration of load balancing in the context of a two-BS system has been used extensively and successfully in previous treatments, e.g.~\cite{yanmaz_jsac_2004,tonguz_ieeetmc_2008,jiang_ton_1996,jeon_jsac_2000}.

Cell 1 has $M_1$ available basic bandwidth units, as referred to in~\cite{deniz_ieeewc_2003} or more commonly referred to as channels, and cell 2 has $M_2$ available channels. Note that channels can also represent WCDMA codes in the context of UMTS. The throughput of every channel in each cell is the same and equal to $R$\,bits/second. We assume that channels are mutually orthogonal and that there is no interference in the set of channels belonging to cells 1 and 2. Each BS emits a signal using omnidirectional antennas and we assume a circular contour signal coverage model, in which full signal strength is received within a certain radius of the BS, and no signal is available beyond that radius, as used in e.g.~\cite{eklundh_ieeetc_1986,tonguz_ieeetmc_2008,wu_ieeejsac_2001,son_ieeetwc_2009}.

\subsection{Node Placement}
\label{sec:node_placement}

Each mobile terminal, referred to as user equipment (UE), remains at fixed positions following an initial UE placement process, with each UE located in one of three separate regions. $N_1$ UEs are in group 1 and have access only to one BS, $N_2$ UEs are in group 2 and have access only to the second BS, and $N_3$ UEs are in group 3 and can potentially access either BS. Such non-homogenous UE placement has been considered, for example in~\cite{jeon_jsac_2000}, which allows for a tractable analysis of the considered system and includes all important groups participating in the load balancing process. The non-homogenous case of UE distribution is the most widespread because UEs are generally distributed non-uniformly over a cellular area. UEs in group 3 are in the region of overlap in coverage between the two cells, also known as the Traffic Transferable Region (TTR)~\cite{tonguz_ieeetmc_2008}. Since only one UE can occupy a channel at a time\footnote{For other, more involved channel assignment procedures including, for example, channel bonding the reader is referred to, e.g.,~\cite{joshi_submitted_2011} where multiple channel assignment problem is studied in an ad hoc scenario.}, a maximum of $M_1+M_2$ UEs can be connected to both cells in the system at any given time. We assume that UEs from groups 1 and 3 are initially registered to cell 1 (serving as the overloaded cell), while UEs in group 2 are initially registered to cell 2.

\subsection{Signal Transmission Model}
\label{sec:signal_transmission_model}

Time is slotted and the minimum time unit is a frame length of $\tau$ seconds. Connections and channel conditions are assumed to remain constant for the duration of a frame, though they will in general vary from frame to frame. We assume that Adaptive Modulation and Coding (AMC) is not used in this framework because it does not aid in evaluating load balancing performance. Please note that AMC is not considered in similar previous works such as~\cite{eklundh_ieeetc_1986,tonguz_ieeetmc_2008,wu_ieeejsac_2001}. For simplicity we also do not consider advanced error control methods such as Hybrid Automatic Repeat Request\footnote{Recent papers~\cite{tomasin_tcom_2009,huang_tvt_2011} are a good source of information on the performance of Hybrid Automatic Repeat Request.}. On the other hand, we do assume that the connection and termination processes for UEs are directly dependent on the channel states experienced between each group of UEs and the BSs they are connected to. We also assume that channel states are binary and independent from slot to slot, much as occurs in~\cite{gotsis_commlet_2009}, and that all of the UEs in each group experience the same channel quality to a given BS. Therefore, in any given time slot a UE is either experiencing a good state with probability $w_{x,y}^{(i)}$, ($x$ denotes the particular pair-wise connection between group $x$ of UEs and associated BS $y$, and $i\in\{d,u\}$ denotes the downlink and uplink, respectively), or a bad state, with probability $1-w_{x,y}^{(i)}$. The value of $w_{x,y}^{(i)}$ is dependent on the distance between UEs in group $x$ and BS $y$, which is denoted as $d_{x,y}$. In our analysis we use the distance as an input to a combined path loss and shadowing propagation model. This serves as an average channel quality consideration for the model instead of the use of channel quality indicators on the uplink per transmitted packet.

\subsection{Prioritization}
\label{sec:prioritization}

Because of the strict boundaries between groups of UEs, we assign priorities on a per group basis. A single, higher priority is given to all UEs in groups 1 and 2 because there is no flexibility to reassign them to a different BS; group 3 UEs can potentially be reassigned and thus given a lower priority. Priorities for UEs are determined on a per time slot basis. A very similar priority model has been used in other treatments of load balancing. For example, in~\cite{jiang_ton_1996,yum_jsac_1993}, newly arriving connections in the non-TTR are given first priority to acquire channels from their serving BSs, while the connections from the TTR are assigned to the remaining channels. Our model allows for the implementation of a wide range of scenarios that require such traffic prioritization. One potential application is for the modeling of networks where load balancing traffic originating from UEs in the TTR has lower priority than non-balanced connections due to several factors including a lower average channel quality~\cite{choi_twc_2007,cruz-perez_twc_2006}, QoS requirements~\cite{li_twc_2005}, non-uniform spatial distribution of traffic classes~\cite{jeon_jsac_2000}, or cell dwell time. Furthermore, it allows for modeling integrated hybrid cellular/WLAN/Ad Hoc networks as discussed in~\cite{song_ieeewc_2009,song_ieeenet_2007}, where non-cellular terminals in the TTR have a lower priority than cellular UEs, and hierarchical cellular systems~\cite{chung_twc_2005}, where members of different tiers have independent priorities. Finally, it enables the modeling of femtocell traffic prioritization, where UEs in groups 1 and 3 are those in the Closed Subscriber Group (CSG)~\cite{choi_globecom_2008}, while UEs in the TTR are neighboring UEs outside of the CSG.

\subsection{Random Access}
\label{sec:random_access}

In the connection process a UE first attempts to connect to the BS it is initially registered to by requesting a connection through a random access channel. We assume a frequency division duplex transmission mode, where control and data traffic are transmitted and received simultaneously. In addition, time division duplexing is considered during the transmission of control packets. Specifically, each UE generates a connection request with probability $p_{x}$. A connection is requested randomly in one of $L_{x}\leq \tau$ non-overlapping, time slotted control resources, unique to group $x\in\{1,2,3\}$ of UEs. In other words, each group has a unique set of sub slots within a frame during which UEs may, but are not required to, request a connection. The random access phase length is equal to slot length $\tau$. Collisions between connection requests from UEs in the same group are possible.

The random access procedure considered in this work shares features of the 3GPP-based cellular network standards, which use the Physical Random Access Channel (PRACH), mapped on a one-to-one basis to the logical random access channel (RACH). RACH uses the S-ALOHA protocol and, in relation to the priorities assumed in this paper, allows the prioritization of connection requests based on Active Service Classes (ASC)~\cite{yang_ieeetwc_2008} which are unique to each UE, and can be adapted by the 3GPP-MAC layer once the UE is in connected mode~\cite[Sec. 2.4.2.6]{laiho_book}. The BS advertises itself to the UEs within range through the broadcast channel using signatures (3GPP release 99, e.g. UMTS), subcarriers (3GPP release 8, e.g. LTE), or time slots, which each ASC can in-turn use for connection requests on RACH. The adaptation of the ASC is performed in the time intervals predefined by the operator. For the purpose of our paper we assume that the BSs collectively, through the Radio Network Controller, map the received signal from every registered UE to an associated ASC.

We assume a zero-persistence protocol, i.e. a collision during a connection request implies that connections are lost, and also UEs do not retry to generate another dependent connection. Due to this assumption a power ramping process, i.e. feedback from the UE to the BS on an unsuccessful connection request~\cite[Sec. II-B]{yang_ieeetwc_2008}, is redundant. To isolate the impact of each group of UEs on collision rates, we assume mutually exclusive RACH resources assigned to each ASC~\cite[Fig. 4]{lin_vtc_2000}. Analysis of PRACH performance in isolation can be found in~\cite{lin_vtc_2000,cooper_tvt_2002}.

\subsection{Data Transfer}
\label{sec:data_transfer}

A connection request is granted during the connection arrangement process if a good channel state occurs between the UE and its associated BS at the moment of the request, and if no collisions occur between multiple requests from different UEs. Once a connection is established, the BS randomly selects a channel and assigns the connected UE to it. The UE then begins to receive downlink data. UEs occupy a time slot with probability $q$, where $1/q=r_{p}/(R\tau)$ is the average connection transfer size and $r_{p}$ is the average packet size given in bits. We assume that the transfer size is at least one time slot long. A connection terminates either when a transmission completes, or when the channel is in a bad state during transmission. 

\subsection{Load Balancing Process}
\label{sec:load_balancing_Process}

In the case of a UE in group 3, if a connection request is successful and there are no resources available in cell 1, we assume that the radio network controller performs load balancing by transferring the call from cell 1 to cell 2\footnote{The network controller maximizes the utilization of resources per single cell to minimize the degradation of performance metrics for its neighboring cells caused by load balancing. As users in group 3 are registered to cell 1 (which is their primary cell), see Section~\ref{sec:node_placement}, and users in group 2 are initially registered to cell 2, load balancing for group 3 users takes place only when all resources in cell 1 are occupied. In the opposite case (when users from group 3 will randomly select a cell for every new connection), the probability of blocking for users in group 2 can potentially increase, causing unnecessary disruption to their quality of service.}. To avoid overloading cell 2 and protecting UEs that are already registered to it, UEs in group 3 can access a maximum of $K$ channels from cell 2, where $0 \le K \le M_2$. UEs in group 3 have access to an additional $K$ shared channels, therefore they have access to a total of $M_1+K$ channels.

Using the nomenclature of~\cite[Sec. 2 and 3]{tonguz_ieeetmc_2008} this load balancing scheme belongs to a class of direct load balancing schemes. It is closest in operation to direct retry~\cite{eklundh_ieeetc_1986}. Since we allow at most $K$ available channels to offload traffic from cell 1 to cell 2 (as in simple borrowing scheme~\cite[Sec. 2]{tonguz_ieeetmc_2008}) we denote our scheme as \emph{direct retry with truncated offloading channel resource pool} (abbreviated as DR$_K$). With $K=M_2$ our scheme reduces to classical direct retry, while for $K=0$ it reduces to a system in which no load balancing takes place.

In our model we do not use a take-back process, i.e. bi-directional load balancing is not considered. That is, once connections from group 3 are offloaded onto cell 2, they remain connected to cell 2 during the transmission despite whether or not resources have been freed in cell 1. In~\cite{wu_ieeemwc_2004} the authors remark that the take-back process, although more fair to cell 2 because it minimizes blocking at cell 2, is not always advantageous to the network due to the high signaling load that accompanies it. Additionally, as in~\cite{tonguz_ieeetmc_2008,wu_ieeejsac_2001}, we do not use queuing, so there is no consideration of a call give-up process~\cite{wu_ieeemwc_2004}. Moreover, we do not allow preemption of connections from the TTR by connections that have access to channels only from their respective BSs.

\section{Analytical Model}
\label{sec:numerical_analysis}

Let $\{{A},{Y}^{(1)},{Y}^{(2)},{C}\}$ denote a state of a Markov system, where ${A}$ denotes the number of resources used by group 1 UEs, ${Y}^{(1)}$ and ${Y}^{(2)}$ denote the number of resources used by group 3 UEs associated with cell 1 and 2, respectively, and ${C}$ denotes the number of resources used by group 2 UEs. Then the steady state probabilities can be denoted as $\pi_{a,b,c,d}\triangleq\Pr({A}=a,{Y}^{(1)}=b,{Y}^{(2)}=c,{C}=d)$. Note that $a+b+c+d\leq \min\{N_1+N_2+N_3,M_1+M_2\}$, $a+b\leq \min\{N_1+N_3,M_1\}$, $c+d\leq\min\{N_2+N_3,M_2\}$, and $b+c\leq \min\{N_3,M_1+K\}$. These conditions govern what states are possible in the transition probability matrix. 

We define the state transition probability as
\begin{align}
r_{a_{t-1},b_{t-1},c_{t-1},d_{t-1}}^{(a_{t},b_{t},c_{t},d_{t})}\triangleq&\Pr({A}_t=a_{t},{Y}_{t}^{(1)}=b_{t},{Y}_{t}^{(2)}=c_{t},\nonumber\\&\qquad{C}_{t}=d_{t}|{A}_{t-1}=a_{t-1},{Y}_{t-1}^
{(1)}=b_{t-1},\nonumber\\&\qquad{Y}_{t-1}^{(2)}=c_{t-1},{C}_{t-1}=d_{t-1}),
\label{eq:rabcd}
\end{align}
where subscripts $t$ and $t-1$ denote the current and the previous time slots, respectively. This allows for computation of the transition probability matrix required to obtain $\pi_{a,b,c,d}$, which is in-turn used to compute the performance metrics of the load balancing-enabled cellular system. In the subsequent sections we describe the process of deriving the transition probability $r_{a_{t-1},b_{t-1},c_{t-1},d_{t-1}}^{(a_{t},b_{t},c_{t},d_{t})}$. We begin by explaining the computation process for the channel quality, and then focus on the derivation of the functions that support (\ref{eq:rabcd}).

\subsection{Derivation of Channel Quality}
\label{sec:outage_probability}

In the downlink, the probability of a UE belonging to group $x\in\{1,2,3\}$ and receiving a good signal from BS $y\in\{1,2\}$, is defined as
\begin{equation}
w_{x,y}^{(\mathsf{d})}\triangleq1-\Pr({J}_{x,y}^{(\mathsf{d})}\leq \gamma_{q}^{(\mathsf{d})})=1-\int_{0}^{\gamma_{q}^{(\mathsf{d})}}p_{{S}_{x,y}^{(\mathsf{d})}}(\gamma,d_{x,y})d\gamma, \!
\label{eq:optage}
\end{equation}
where $\gamma_{q}^{(\mathsf{d})}$ is the signal reception threshold for the downlink, expressed as the minimum required received power of the received signal ${J}_{x,y}^{(\mathsf{d})}$, and $p_{{S}_{x,y}^{(\mathsf{d})}}(\gamma,d_{x,y})$ is the distribution of the signal $\gamma$ received at group $x$, which is at a distance of $d_{x,y}$ from BS $y$. As an example, we consider an environment with path loss and shadowing, where $w_{x,y}^{(\mathsf{d})}$ is expressed in~\cite[Eq. 2.52]{goldsmith_book} as
\begin{equation}
w_{x,y}^{(\mathsf{d})}\!=\!Q\!\!\left(\frac{\gamma_{q}^{(\mathsf{d})}\!\!-\!\!P_{t}^{(\mathsf{d})}\!\!-\!\!10\log_{10}W^{(\mathsf{d})}\!\!+\!\!10\delta\log_{10}\frac{d_{x,y}}{d_{0,x,y}^{(\mathsf{d})}}}{\sigma_{\Psi}}\right),
\label{eq:wxy}
\end{equation}
where $W^{(\mathsf{d})}$ is a unit-less constant dependent on both the antenna characteristics and an average channel attenuation, and for $W^{(\mathsf{d})}<1$ approximating empirical measurements and assuming omnidirectional antennas is given as~\cite[Eq. 2.41]{goldsmith_book},
\begin{equation}
W^{(\mathsf{d})}\text{ dB }=20\log_{10}\frac{\lambda}{4\pi d_{0,x,y}^{\mathsf{(d)}}},
\label{eq:w_def} 
\end{equation}
where $\lambda$ is the wavelength of the carrier frequency; $P_{t}^{(\mathsf{d})}$ is the BS transmitted power (which is assumed to be the same for both base stations, $d_{x,y}$ is the distance of the UE in group $x$ located farthest from BS $y$; $d_{0,x,y}^{(\mathsf{d})}$ is the reference distance for the BS antenna far-field; $\sigma_{\Psi}$ is the log-normal shadowing variance given in dB; $\delta$ is the path loss exponent; and the $Q$ function is defined in the usual manner as
\begin{equation}
Q(z)\triangleq\int_{z}^{\infty}\frac{1}{\sqrt{2\pi}}e^{-u^{2}/2}du.
\end{equation}
In the uplink, the probability that a good signal is received by BS $y$ from a UE in group $x$ is $w_{x,y}^{(\mathsf{u})}$, and is expressed in the same manner as equations (\ref{eq:optage}) and (\ref{eq:wxy}), by replacing all variables having superscript $(\mathsf{d})$ with $(\mathsf{u})$, where $P_{t}^{(\mathsf{u})}$ denotes the UE transmission power; $W^{(\mathsf{u})}$ denotes the constant for the UE antenna (which again can be calculated in the same manner as $W^{(\mathsf{d})}$ using (\ref{eq:w_def})); $d_{0,x,y}^{(\mathsf{u})}$ is the reference distance for the UE antenna far-field; and $\gamma_{q}^{(\mathsf{u})}$ is the signal reception threshold for the uplink. The downlink and uplink channel quality information governs the success rate of the connection admission process, as well as the duration of a downlink transmission.

\subsection{Derivation of Connection Arrangement Probability}
\label{sec:connection_arrangement}

An important feature of the model is the consideration of connection admission in the load-balancing process. This process is a function of the total number of UEs, the number of UEs receiving data from their respective serving BSs, the pairwise channel quality between the UEs and its serving BSs, and the underlying random access algorithm. The probability that $j$ new connections have successfully requested downlink data given $i_{t-1}\in\{a_{t-1},b_{t-1},c_{t-1},d_{t-1}\}$ currently active connections from group $x$, with a random access channel consisting of $L_x$ time slots is
\begin{equation}
S_{i_{t-1}}^{(j)}\triangleq
\begin{cases}
\zeta_{i_{t-1},x,y}^{(i_{t})}, & j>0,\\
1-{w}_{x,y}^{(\mathsf{u})}+\zeta_{i_{t-1},x,y}^{(1)}, & j=0,\\
0, & \text{otherwise},
\end{cases}
\label{eq:sij}
\end{equation}
where
\begin{align}
\zeta_{i_{t-1},x,y}^{(j)}\triangleq&
\sum_{k=j}^{N_{x}-i_{t-1}}\binom{N_{x}-i_{t-1}}{k}p_{x}^{k}(1-p_{x})^{N_{x}-i_{t-1}-k}\beta_{k}^{(j)}\nonumber\\&\qquad\times w_{x,y}^{(\mathsf{u})};
\end{align}
$p_{x}$ is the probability of a connection request by an individual UE in group $x$; and $\beta_{k}^{(j)}$ is the probability that among ${k}$ UEs requesting a connection, $j$ were successful in obtaining a resource. Note that the reference to $x,y$ in $S_{i_{t-1}}^{(j)}$ is omitted due to space constraints, keeping in mind that for $a_{t},a_{t-1}$ $x=1$, $y=1$, for $b_{t},b_{t-1}$ $x=3$, $y=1$, for $c_{t},c_{t-1}$ $x=3$, $y=2$, and for $d_{t},d_{t-1}$ $x=2$, $y=2$.

For consistency with cellular networks such as 3GPP, we consider a PRACH-like control channel, for which $\beta_{k}^{(j)}$ can be described in the manner of~\cite[Eq. (3)]{zhou_ieeetwc_2008}
\begin{equation}
\beta_{k}^{(j)}=\sum_{m=j}^{\min(k,L_x)}\frac{(-1)^{m+j}(L_{x}-m)^{k-m}k!}{(m-j)!(L_{x}-m)!(k-m)!n!}.
\label{eq:prnk}
\end{equation}
Note that depending on the assumption of how collisions are resolved, different definitions of $\beta_{k}^{(j)}$ in (\ref{eq:prnk}) can be applied when calculating the connection arrangement probability according to~(\ref{eq:sij}).

\subsection{Derivation of Connection Termination Probability}
\label{sec:connection_termination}

Once a UE successfully requests a connection from the serving BS, a downlink transmission is started provided that at least one free channel is available for the UE. The connection terminates when the BS finishes transmitting data to the UE or when the downlink signal received by the UE is in outage. The probability that $j$ connections from $i_{t-1}$ active connections at group $x$ terminate is
\begin{equation}
T_{i_{t-1}}^{(j)}\triangleq\binom{i_{t-1}}{j}l_{x,y}^{j}(1-l_{x,y})^{i_{t-1}-j},
\label{eq:tij}
\end{equation}
where $l_{x,y}=1-w_{x,y}^{(\mathsf{d})}+w_{x,y}^{(\mathsf{d})}q$ is the inverse of the average packet length, accounting for truncation of some packets due to a bad channel quality. Again, the indices $x,y$ have been omitted for notational simplicity in the symbol $T_{i_{t-1}}^{(j)}$, assuming that the same relationship between $x,y$ and $j,i_{t-1}$ as given in Section~\ref{sec:connection_arrangement} holds.

\subsection{Derivation of the State Transition Probability}
\label{sec:derivation_steady_state}

Using the definitions of the arrangement and termination probabilities, expressed in (\ref{eq:sij}) and (\ref{eq:tij}), respectively, we can finally introduce the state transition probabilities for the complete model. The transition probability is constructed using the termination and arrangement probability definitions and the respective relationship between the variables of these two definitions (which are dependent on the start and end states of the transition). Due to the complexity of the derivation we begin with a highly simplified example.

\subsubsection{State Transition Probabilities for a Single UE Group}

To facilitate understanding the derivation of the complete state transition probabilities, we first consider a network in which no load balancing occurs and all of the UEs are in group 1, such that $N_1>0$ and $N_2=N_3=0$. The state of the Markov chain then simplifies to $\{A,0,0,0\}$ and the transition probability becomes $r_{a_{t-1},0,0,0}^{(a_{t},0,0,0)}$, where
\begin{equation}
r_{a_{t-1},0,0,0}^{(a_{t},0,0,0)}\!\!=\!\!
\begin{cases}
\sum_{i=0}^{a_{t}}T_{a_{t-1}}^{(i+a_{t-1}-a_{t})}S_{a_{t-1}}^{(i)}, & \!\!\!\!\begin{split}a_{t-1}\geq a_{t}, \\a_{t}< M_{1};\end{split}\\
\sum_{i=0}^{a_{t}}T_{a_{t-1}}^{(i)}S_{a_{t-1}}^{(i+a_{t}-a_{t-1})}, & \!\!\!\!\begin{split}a_{t-1}<a_{t},\\ a_{t}<M_{1};\end{split}\\
\sum_{i=0}^{a_{t}}T_{a_{t-1}}^{(i)}S_{a_{t-1}}^{(i+a_{t}-a_{t-1})}\\\quad+\sum_{i=0}^{a_{t}}T_{a_{t-1}}^{(i)}\\\quad\times\sum_{j=M_{1}}^{N_{1}}S_{a_{t-1}}^{(i+j-a_{t-1})}, & \!\!\!\!\begin{split}a_{t-1}\leq a_{t},\\ a_{t}=M_{1},\end{split}
\label{eq:rat-1at}
\end{cases}
\end{equation}
In (\ref{eq:rat-1at}) the case $a_{t-1}\geq a_{t}$, $a_{t}< M_{1}$ denotes the transition from a higher to a lower channel occupancy, subject to the constraint that the number of channels occupied in the end state must be less than the total BS capacity. The number of terminating UEs is set to compensate for the UEs that generate successful connections. The case $a_{t-1}<a_{t}$, $a_{t}< M_{1}$ denotes the transition from a lower to a higher channel occupancy (given, again, that the number of occupied channels is less than the total BS capacity). In this case UEs from group 1 need to generate exactly as many connections as given by the end state, not forgetting to generate connections in order to compensate for the total number of terminations. Lastly, for the case of $a_{t-1}\leq a_{t}$, $a_{t}=M_{1}$ the end state equals the total channel capacity. The first term in the definition of this transition probability includes exactly the number of connections needed to reach the end state, once again compensating for terminations. The second term accounts for all successful connections generated that exceed those needed to reach the end state, which will not be admitted.

\subsubsection{General Solution for the State Transition Probabilities}

Due to the complexity of the general solution, the main analytical equations describing the construction of the four-dimensional transitional probability matrix are presented in the Appendix~\ref{sec:appendix}.

\subsection{Performance Metrics}

Given the complete description of the system, we are able to derive important metrics that would describe the efficiency of the load balancing process involving connection admission. While there are many performance metrics that can be extracted given the above framework, we focus on three primary metrics: (i) the blocking probability, which describes the probability that at least one UE which requests a connection from a particular group will be denied access to a channel, (ii) the channel utilization, which expresses the fraction of the available channels are being used, and (iii) the collision probability on a control channel, which provides the probability that at least one requesting connection will be lost due to a collision with another UE.

\subsubsection{Blocking Probability}

As used here, blocking occurs when at least one UE requests a new connection, but cannot be admitted to any BS due to lack of available channels. Since each group has access to a different number of channels and can follow a different connection strategy, it is necessary to define separate blocking probability metrics for UEs in groups 1 and 2, as contrasted with UEs in group 3. For groups 1 and 2, the blocking probability is defined according to
\begin{align}
B^{(z)}=&\sum_{a,b,c,d}\sum_{i=0}^{x}\sum_{j=M_{v}-x-y+1}^{N_{z}}\sum_{k=0}^{y}\pi_{a,b,c,d}T_{x}^{(i)}T_{y}^{(k)}\nonumber\\&\qquad\times S_{x}^{(i+j+k)},
\end{align}
where for $z=1$ $x=a$, $y=b$, $v=1$ and for $z=2$ $x=d$, $y=c$, $v=2$. For group 3 UEs, the blocking probability is given as
\begin{align}
B^{(3)}=&\sum_{a,b,c,d}\sum_{i_a=0}^{a}\sum_{j_a=0}^{N_1}\sum_{i_d=0}^{d}\sum_{j_d=0}^{N_3}\sum_{i=0}^{b}\sum_{j=0}^{c}\sum_{k=\phi}^{N_2}
\pi_{a,b,c,d}T_{a}^{(i_a)}S_{a}^{(j_a)}\nonumber\\&\qquad \times T_{d}^{(j_d)}S_{d}^{(j_d)}T_{b}^{(i)}T_{c}^{(j)}S_{b+c}^{(i+j+k)},
\end{align}
where $\phi$ is defined separately for $K<M_2$ and $K=M_2$. For $K=M_2$ $\phi=M_{1}+M_{2}-a-b-c-d-g_a-g_d+i_a+i_d+1$, where
\begin{equation}
g_a=
\begin{cases}
M_{1}-a\\\quad+i_a-b+i, & j_a>M_{1}-a+i_a-b+i,\\
j_a, & \text{otherwise},
\end{cases}
\label{eq:g_a}
\end{equation}
and $g_d$ is defined as~(\ref{eq:g_a}) replacing $M_{1}$ with $M_{2}$, $a$ with $c$, $i_a$ with $i_d$, $b$ with $d$ and $i$ with $j$. For $K<M_2$ $\phi=M_{1}-g_a-b-j_a+i_a+i+g_c+1$, where $g_a$ is defined as (\ref{eq:g_a}) and
\begin{equation}
g_c=
\begin{cases}
\max\{0,g_e\}, & g_e < K,\\
K-c+j, & c-j<K \text{~and~} g_e \geq K,\\
0, & \text{otherwise},
\end{cases}
\label{eq:g_c}
\end{equation}
where $g_e=M_2-c-d+i_d+j-j_d$.

We briefly explain the above equations. The derivation of the blocking probability for group 3 UEs is more complicated than for those UEs in groups 1 and 2 because this group can access channels from both cells. Therefore, the number of generations for group 3 UEs that leads to blocking has to account for the terminations within the same group, and also for the possible changes in the number of connections of UEs in groups 1 and 2.

The blocking for group 3 UEs can be analyzed in two separate cases. The first case accounts for the number of generations needed to occupy all the channels in cell 1 ($K=M_2$), while the second case accounts for the number of generations needed to occupy all available channels in cell 2 ($K<M_2$). The first case is simpler to analyze because group 3 UEs need only generate as many connections as there are available resources on cell 1.

When $K < M_{2}$, group 3 UEs can only access a maximum of $K$ channels on cell 2. Therefore, extra conditions are added for the situation in which group 3 UEs are blocked when exceeding $K$ connections in cell 2. If there are less than $K$ available free channels after terminations of group 3 UEs connected to cell 2, the function $\max\{0, g_e\}$ represents the number of connections to cause blocking by generating the exact number of connections to occupy all available channels. The $\max\{0, g_e\}$ function is used to lower bound the necessary number of connections for blocking. This is because the number of connections on cell 2, in general, is not restricted to $K$ and can thus have more than $K$ current occupancies resulting in a possibly negative value for $g_e$. On the other hand, if there are more than $K$ available channels after terminations, exactly $K$ channels are used by group 3 UEs to cause blocking.

\subsubsection{Channel Utilization and Total System Throughput}

The overall channel utilization is
\begin{equation}
U=\frac{1}{M_{1}+M_{2}}\sum_{a,b,c,d}(a+b+c+d)\pi_{a,b,c,d},
\label{eq:u}
\end{equation}
which refers to the fraction of the collective capacity that has been used by all UEs in all groups. The average total system throughput is obtained by multiplying (\ref{eq:u}) by $R$.

\subsubsection{Collision Probability}

Because the model uses a random access channel for connection requests, it is
necessary to compute the collision probability of the system. The collision probability
for UEs in group $x$ is:
\begin{align}
D^{(x)}=\sum_{a,b,c,d}\sum_{k=0}^{N_{x}-a}\sum_{j=0}^{K}\beta_{k}^{(j)}I_{k-j}^{(1)}\pi_{a,b,c,d}\binom{N_{x}-\eta}{k}\nonumber\\\quad\times p^{k}(1-p)^{N_{x}-\eta-k}w_{x,y}^{(\mathsf{u})},
\end{align}
where $I_{i}^{(j)}=1$ when $i \geq j$ and $I_{i}^{(j)}=0$, otherwise, $\eta=\{a,d\}$ for group $x=\{1,2\}$, respectively, and $\eta=b+c$ for group $x=3$.

\subsection{A Special Case: Load Balancing with Opportunistic Spectrum Access}
\label{sec:osa_balancing}

In this section we illustrate the use of the proposed system to analyze other, more complex, cellular setups. As an example, we present a case where the connections can be offloaded to a neighboring cell belonging to a separate network (such as that of another provider when there is neither shared signaling between cells nor a central controller to coordinate the distribution of connections from one system to another). In the proposed system, users registered to one cell must detect the availability of free channels in the neighboring cell by means of passive channel observations, i.e. spectrum sensing. Such a system can be referred to as Opportunistic Spectrum Access (OSA)-enabled load balancing~\cite{buddhikot_dyspan_2010}.

We note that very little work has been done to consider the system-level performance of OSA-enabled cellular networks. A group of papers, e.g.~\cite{ma_tcomm_2010,xiang_systj_2010}, analyzes sharing mechanisms of radio resources based on OSA, focusing on the physical characteristics, while abstracting higher layers. These papers provide insights on how OSA can be beneficial in a cellular system, but do not address the specific system performance and tradeoffs involved---specifically, the traffic characteristics of primary and secondary cells and the connection admission process during the random access phase. While many papers have analyzed the performance of OSA networks, e.g.~\cite{Park_arxiv_2009,almahdi_commlet_2009}, the connection between OSA and its use in cellular load balancing has not been deeply addressed.

\subsubsection{Extensions to the System Model}
\label{sec:extensions_osa_system_model}

The following changes need to be made to the system model.

\paragraph{Spectrum Sensing}

First, due to spectrum sensing assumption, each user in cell 1 is equipped with a spectrum sensor in order to detect whether any of the channels on cell 2 is occupied. It is assumed that $\tau_{s}<\tau$ seconds are needed for spectrum sensing within each time slot. Channel availability detection is assumed to be imperfect. We denote $\mu_f$ as the probability of false alarm, i.e. that users in cell 1 will declare that a channel is occupied in cell 2 when in fact it is not. If users from group 3 attempt to access a channel detected as occupied by users from cell 2, the users in group 3 are blocked from connecting to that channel. The probability of mis-detecting the presence of a vacant channel in cell 2 is denoted as $\mu_m$. If a mis-detection occurs, it is assumed that the users from group 3 will occupy the channel at cell 2 and successfully be able to receive downlink data despite the presence of the licensed users in cell 2. We assume that as long as the users in cell 1 fulfills the minimum requirement on detection probability, $1-\mu_m$, it does not violate QoS requirements of the users in cell 2. We also assume that the coding scheme of the users in cell 1 can be designed to compensate for the increased level of interference in the event of a mis-detection. An existing connection of any user from cell 1 is preempted by users in cell 2 that accesses the same channel on cell 2. Once user from group 3 is preempted its connection is dropped.

\paragraph{Connection Termination}

Cell 2 users receive downlink data from the BS using channels assigned to cell 2 until the earlier of (i) the termination of the connection after successful transmission of all data, or (ii) preemption by users in cell 2 (if cell 2 users occupy the same channel). For the sake of brevity, we assume that perfect channel conditions are experienced in the system, i.e. $w_{x,y}^{(i)}, i\in\{\mathsf{u},\mathsf{d}\}$. Due to spectrum sensing phase the average packet length is now $1/q=r_p/(R\left(\tau-\tau_s\right))$.

\paragraph{Load Balancing Scheme}

The same load balancing process is used, DR$_K$, however we assume $K=M_2$ throughout the analysis of OSA system, as control over access to channels used for load balancing is governed by the spectrum sensing quality, not by size of the channel pool accessible to cell 1. For the sake of brevity we assume that there are no users in the group 1 and 2. Instead, we assume that users in cell 2 occupy any channel on cell 2, with a geometrically distributed average slot occupancy probability of $q_p$. As cell 2 does not take part in load balancing process this choice is well motivated.

\subsubsection{Analytical Model}
\label{sec:analysis_osa}

As in case of general load balancing system, we construct a Markov chain to derive performance metrics. We reuse the notation from the general load balancing system. Let a state of a Markov system be given as $\{Y^{(1)},Y^{(2)},{C}\}$, where $Y^{(1)}$ denotes the number of users from group 3 connected to cell 1, $Y^{(2)}$ denotes the number of users from group 3 connected to cell 2, and ${C}$ denotes the number of users of cell 2 connected to cell 2. We denote the steady state probabilities as $\pi_{b,c,d}\triangleq\Pr({Y^{(1)}}=b,{Y^{(2)}}=c,{C}=d)$. We also define the state transition probability $r_{b_{t-1},c_{t-1},d_{t-1}}^{(b_{t},c_{t},d_{t})}=\Pr(Y_t^{(1)}=b_{t},Y_t^{(2)}=c_{t},{C}=d_{t}|Y_{t-1}^{(1)}=b_{t-1},Y_{t-1}^{(2)}=c_{t-1},{C}_{t-1}=d_{t-1})$, where subscripts $t$ and $t-1$ denote the current and the previous time slots, respectively. The transition probabilities allow for the computation of the transition probability matrix which is subsequently used to compute the steady-state distribution, which is denoted by $\pi(b,c,d)$.

The transition probabilities are governed by arrangement probability, defined as (\ref{eq:sij}), termination probability, defined as (\ref{eq:tij}), and preemption probability defined as~\cite[Sec. IV-C1]{Park_arxiv_2009}
\begin{equation}
P_{x,y}^{(i)}=\left({x+i\atop i}\right)\left({M_{2}-x-i\atop y-i}\right)q_{p}^{y}\left(1-q_{p}\right)^{M_{2}-y},
\label{eq:Pxyi}
\end{equation}
where $x$ is the current number of group 3 users connections on cell 2, $y$ is the current number of connections occupied by users of cell 2 and $i$ is the number of incoming group 3 users connection generations.

Having these three equations, we can derive the set of equations that describe the transition probability matrix for the general solution. They are presented in the Appendix~\ref{sec:appendix_osa}.

\subsubsection{Performance Metrics}
\label{sec:performance_metrics_osa}

We can define two important performance metrics for the model.

\paragraph{Group 3 Users Throughput}
\label{sec:su_throughput}

The Group 3 users throughput is defined as
\begin{equation}
Z=\frac{\tau-\tau_s}{\tau}R\sum_{b,c,d}\left(b+c\right)\pi\left(b,c,d\right).
\label{eq:u_osa}
\end{equation}

\paragraph{Collision Probability Between Cell 1 and cell 2 Users}
\label{sec:pu_su_collision}

A collision occurs when users in group 3 occupy the same channel as users in cell 2 after mis-detecting their presence within a time slot. The probability that there is a mis-detected channel is given by $x_q=q_{p}\mu_m$. The probability that there are $y$ collisions in a time slot is denoted as $\kappa_y$ and is described as follows. If $y \ge b$
\begin{align}
\kappa_y&=\sum_{a,b,c}\pi\left(a,b,c\right)\sum_{m=y}^{M_{2}}\binom{M_{2}}{m}x_{q}^{m}\left(1-x_{q}\right)^{M_{2}-m}\nonumber\\&\times\left(\sum_{k=0}^{a}\sum_{l=0}^{b}\sum_{r=0}^{M_{2}-m}T_{a}^{k}T_{b}^{l}S_{a+b}^{(y-b+r+k+M_{1}-a+l)}\frac{\binom{m}{y}\binom{M_{2-m}}{r}}{\binom{M_{2}}{y+r}}\right.\nonumber\\&\left.+{^{(0)}I_{y}^{(m)}}\sum_{k=0}^{a}\sum_{l=0}^{b}\sum_{r=M_{2}-b+M_{1}-a+1}^{N_{2}}\!\!\!\!\!\!\!\!T_{a}^{k}T_{b}^{l}S_{a+b}^{(k+l+r)}\right).
\label{eq:C_y1}
\end{align}
If the number of collisions is greater than or equal to the current number of occupancies of group 3 users on cell 2, the number of group 3 users connection generations on cell 2 are increased in order to meet the number of collisions. When the number of group 3 users connection generations on cell 2 exceeds the desired number of collisions, a hypergeometric term is used to calculate the exact probability of having $y$ collisions. When $y=m$ an excess term is needed for additional group 3 users connection generations that cannot be accommodated. On the other hand, if $y < b$
\begin{align}
\kappa_y&=\sum_{a,b,c}\pi\left(a,b,c\right)\sum_{m=y}^{M_{2}}\binom{M_{2}}{m}x_{q}^{m}\left(1-x_{q}\right)^{M_{2}-m}\left(\sum_{k=0}^{a}\sum_{l=0}^{b}\right.\nonumber\\&\left.\times\sum_{r=\max\left(y-b+l,0\right)}^{M_{2}-b+l}T_{a}^{k}T_{b}^{l}S_{a+b}^{(r+M_{1}-a+k)}\frac{\binom{m}{y}\binom{M_{2}-m}{b+r-k-y}}{\binom{M_{2}}{b+r-k}}\right.\nonumber\\&\left.+{^{(0)}I_{y}^{(m)}}\sum_{k=0}^{a}\sum_{l=0}^{b}\sum_{r=M_{2}-b+M_{1}-a+1}^{N_{2}}\!\!\!\!\!\!\!\!T_{a}^{k}T_{b}^{l}S_{a+b}^{(k+l+r)}\right).
\label{eq:C_y2}
\end{align}
If the number of collisions is strictly less than the number of occupancies on cell 2 then additional group 3 users connection generations are unnecessary. However, if additional group 3 users terminations occur, group 3 users generations are needed to ensure at least an equal number of collisions. In turn, the average number of collisions is defined as $\kappa_{a}=\sum_{y=0}^{M_{2}}y\kappa_{y}$.

\section{Results}
\label{sec:numerical_results}

Since our model incorporates a very large number of parameters, in the interest of clarity and brevity we focus our study on certain scenarios that are the most important in the context of our model. First, we present results that demonstrate the impact of a varying channel quality on the load balancing efficiency. Second, we examine the influence of the random access phase on load balancing efficiency. Third, we provide insight on the optimal channel sharing policy between BS 1 and BS 2. Fourth, we present results on load balancing in the OSA context in Section~\ref{sec:osa_results}. And finally, in Section~\ref{sec:multicell} we present insight on how to extend our analytical model to a multi-cell scenario.

To confirm the correctness of the analytical model, we created a simulation environment for verifying the analytical results. The results in Section~\ref{sec:cqi}, Section~\ref{ses:varying_p},  Section~\ref{sec:osa_results} and Section~\ref{sec:multicell} obtained using both the analytical and simulation approaches to confirm correctness, while those in Section~\ref{sec:varying_k} are obtained using simulation.

Note, that our model is related to~\cite{eklundh_ieeetc_1986,yanmaz_jsac_2004,wu_ieeejsac_2001}. However, the exact comparison of our model with~\cite{eklundh_ieeetc_1986,yanmaz_jsac_2004,wu_ieeejsac_2001} is impossible, due to the following differences: (i) our model and those of~\cite{eklundh_ieeetc_1986,yanmaz_jsac_2004,wu_ieeejsac_2001} consider a system where users are uniquely identified and treated as a single group, respectively; (ii) our model, in contrary to~\cite{eklundh_ieeetc_1986,yanmaz_jsac_2004,wu_ieeejsac_2001} considers a more involved connection allocation process where, in addition to channel allocation (considered in~\cite{eklundh_ieeetc_1986,yanmaz_jsac_2004,wu_ieeejsac_2001} only), connection admission through a random access channel is analyzed; (iii) model of~\cite{yanmaz_jsac_2004,wu_ieeejsac_2001} considers queuing, while our model does not (for tractability reasons).

\subsection{Impact of Channel Quality on Load Balancing Process}
\label{sec:cqi}

In this simulation, we model, among others, the call admission, termination, and load balancing processes exactly as described in our system model. As an example, we consider a scenario in which two identical cells are positioned such that they form a small TTR. For simplicity, we assume that $N_x=L_x$, where $x\in\{1,2,3\}$ and $M_1=M_2=K$. This particular analysis represents the effect of an increasing $w_{3,1}^{(i)}$ on the overall system-wide channel utilization, while setting $w_{1,1}^{(i)}=w_{3,2}^{(i)}=w_{2,2}^{(i)}=0.806$ for all $i\in\{u,d\}$, assuming reciprocal uplink and downlink conditions. This is equivalent to varying $d_{3,1}$ from a location that is out of range to being right next to BS 1, and setting $d_{1,1}=d_{3,2}=d_{2,2}=30$\,m. We use the pathloss model with the following parameters: $\gamma_{q}^{(i)}=-85$\,dBm, $P_{t}^{(i)}=30$\,dBm, $W^{(i)}=7.01(10)^{-4}$, $\sigma_{\Psi}=3.65$\,dB, $d_{0,x,y}^{(i)}=1$\,m and lastly $\delta=4.76$. Furthermore, we assume an average channel capacity of $R=250$\,kbps per channel with an average packet size $r_p=1$\,kB and $\tau=8$\,ms slot length. This yields an average packet length of $31.25$ time slots, or equivalently 32\,ms and probability of time slot occupancy of $q=0.25$. The channel throughput represents a typical value used in radio access network planning calculations~\cite[Table 8.17]{holma_book}. The packet size represents a realistic packet length sent over the Internet~\cite{williamson_comput_2001}, where packets are distributed between a minimum value of 40\,B (Transport Control Protocol acknowledgement packet) and a Maximum Transmission Unit, which for IPv6 equals 1.268\,kB, for IEEE 802.3 equals 1.492\,kB, for Ethernet II equals 1.5 kB, and for IEEE 802.11 equals 2.272\,kB.
\begin{figure*}
\centering
\subfigure[]{\includegraphics[width=0.49\columnwidth]{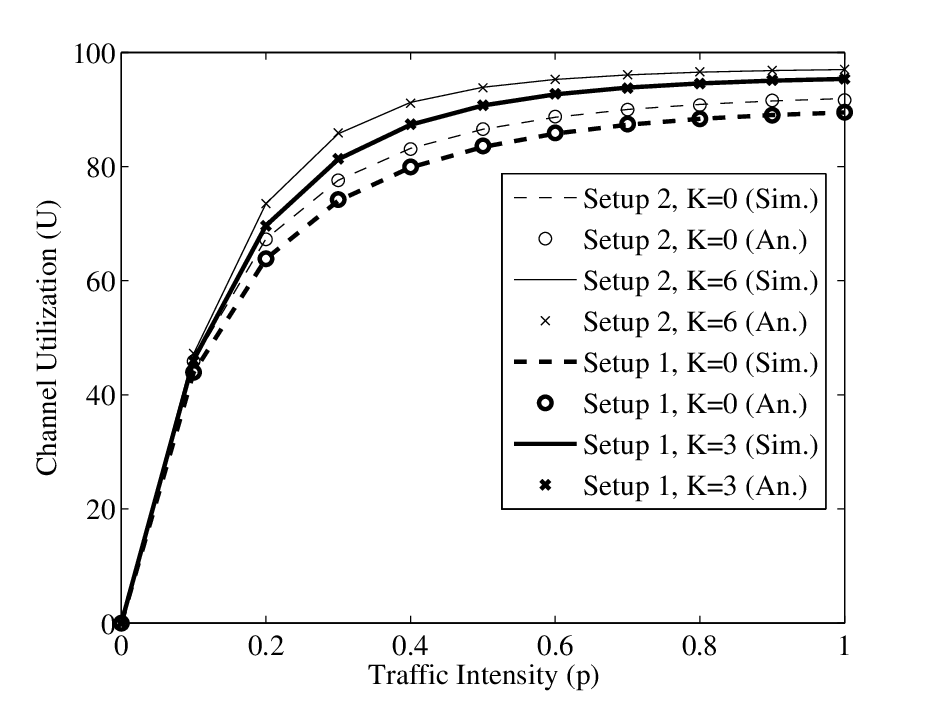}\label{fig:chan_util}}
\subfigure[]{\includegraphics[width=0.49\columnwidth]{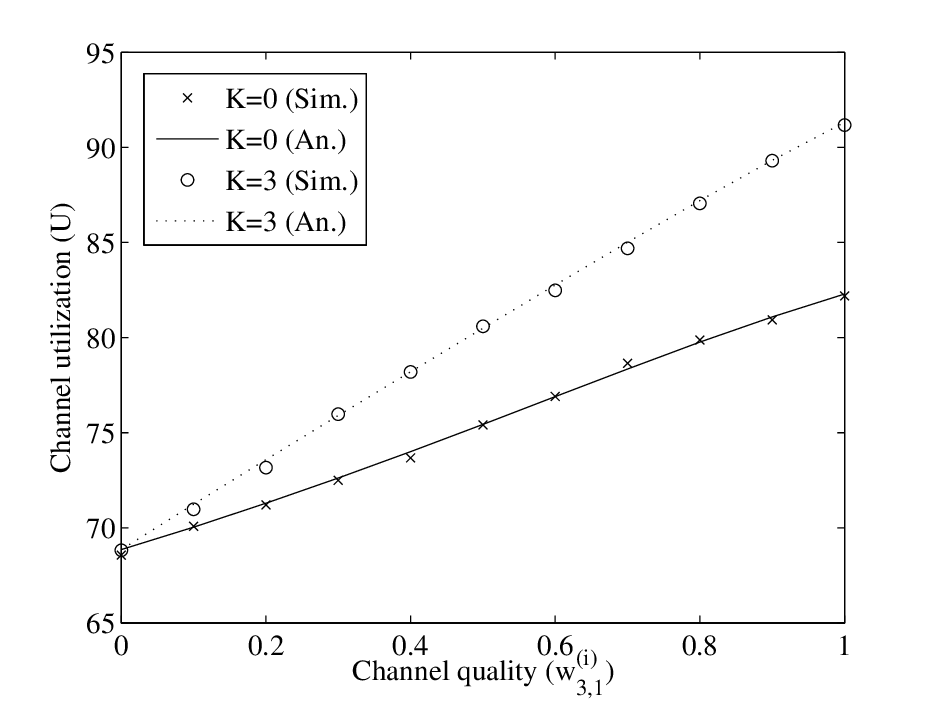}\label{fig:chan_util_w31}}
\caption{Impact of $K$ on channel utilization: (a) The channel utilization as a function of traffic intensity $p$. Two extremes of shared channels, i.e. $K=0$ (no load balancing) and $K=M_2$ (all of cell 2's channels used in load balancing) are shown. Furthermore, two network setups are considered (a) Setup 1: $L_x=N_x=6$, $M_1=M_2=3$, and (b) Setup 2: $L_x=N_x=12$, $M_1=M_2=6$, where $x\in\{1,2,3\}$; the other parameters of the network are common for both cases and described in Section~\ref{sec:cqi}. The figure shows good agreement between the results from the analytical model and from simulation; (b) The channel utilization, represented as a percentage on the vertical axis, as a function of the channel quality between group 3 UEs and BS 1, $w_{3,1}^{(i)}$ for two extreme values of $K$. As $w_{3,1}^{(i)}$ improves, more group 3 UEs generate successful connections to cell 1 resulting in more UEs that connect to BS 1 and consequently are offloaded onto cell 2, resulting in an overall increase in channel utilization.}
\end{figure*}

In order to determine the best performance of the load-balancing scheme, represented by the proposed DR$_K$ algorithm (in the context of the considered system mode and chosen parameter values), for a varying channel quality we determine the level of traffic intensity which results in maximum channel utilization. Fig.~\ref{fig:chan_util} expresses the channel utilization as a function of increasing traffic intensity $p_{1}=p_{2}=p_{3}=p$ for two extreme $K$ values, i.e. $K=0$ (when no load-balancing is used) and $K=M_2$ (all of cell 2's channels may be used for load balancing) considering two network setups: setup 1 with $L_x=N_x=6$, $M_1=M_2=3$, and setup 2 with $L_x=N_x=12$, $M_1=M_2=6$. As expected, the channel utilization increases with more traffic intensity because an increase in $p$ results in more frequent connection requests from UEs in all groups leading to a higher probability of successful connections. Moreover, the larger the number of channels and users in both cells, the larger the channel utilization--for both cases considering $K=0$ and $K=M_2$. The increase in channel utilization tails off as the system reaches saturation, i.e. close to 100\% channel utilization. Similarly, an increase in $K$ results in a higher channel utilization as more UEs from group 3, that are blocked from cell 1, are offloaded onto cell 2 where they have access to an additional $K$ channels. We observe that there is a decreasing rate of gain in channel utilization with an increase in the number of shared channels $K$. As Fig.~\ref{fig:chan_util} shows, a traffic intensity of $p=0.4$ is the point at which the system begins to operate in saturation, i.e. the relative difference between channel utilizations for $K=M_2$ and $K=0$ remains relatively constant thereafter for both network setups. With the knowledge of decreasing gains in channel utilization with increasing $K$, there may exist an intermediate value of $K$ that not only leads to an improvement in total channel utilization, but also maximizes improvement with respect to the overall UE experience. This value of $K$ is explored in Section~\ref{sec:varying_k}. In the current section we continue our investigation using $p=0.4$ and explore the impacts of channel quality on performance.

Fig.~\ref{fig:chan_util_w31} illustrates an increase in channel utilization with an increase in $w_{3,1}^{(i)}$ for two extreme values of $K$. Increasing $w_{3,1}^{(i)}$ results in group 3 UEs having more successful requests for receiving downlink transmissions because the average channel quality, in which requests are granted for group 3 UEs, improves. Therefore, ignoring the channel effects by assuming perfect channel conditions (also done by setting $w_{3,1}^{(i)}=1$ in our model) in the analysis of load-balancing schemes, even for one particular group of UEs, produces a non-trivial difference in the channel utilization and leads to an exaggerated improvement in performance due to load balancing. By selecting a reasonable scheme to determine the channel quality, as presented in Section~\ref{sec:outage_probability}, we are able to provide a more realistic evaluation of the improvements of load balancing. Note that the average channel utilization significantly increases as more channels can be borrowed from BS 2. When $w_{3,1}^{(i)}$ increases, the difference in channel utilization between cases $K=0$ and $K=3$ becomes more profound. This proves that with low channel quality system-wide improvement from load balancing might not be as significant as in the case of perfect channel conditions.
\begin{figure*}
\centering
\subfigure[$K=0$]{\includegraphics[width=0.49\columnwidth]{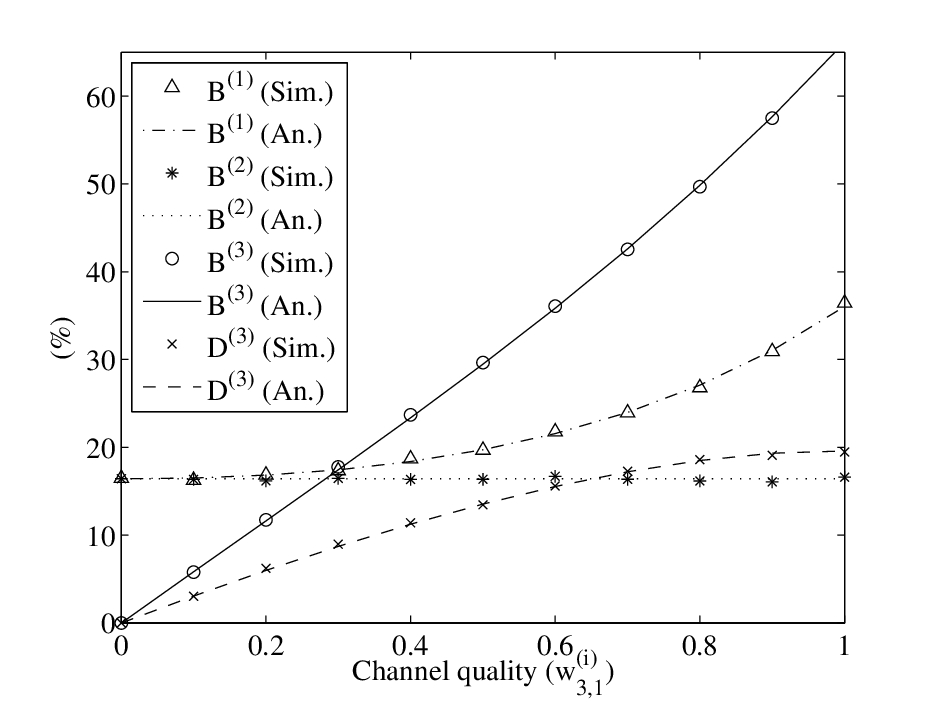}\label{fig:incr_w31_K=0}}
\subfigure[$K=1$]{\includegraphics[width=0.49\columnwidth]{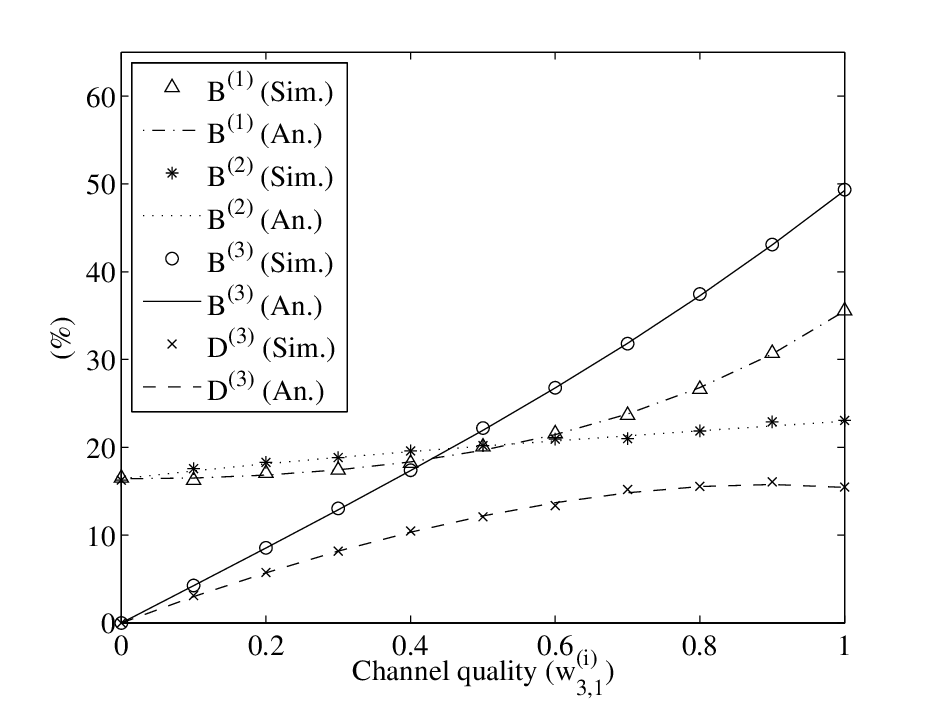}\label{fig:incr_w31_K=1}}\\
\subfigure[$K=2$]{\includegraphics[width=0.49\columnwidth]{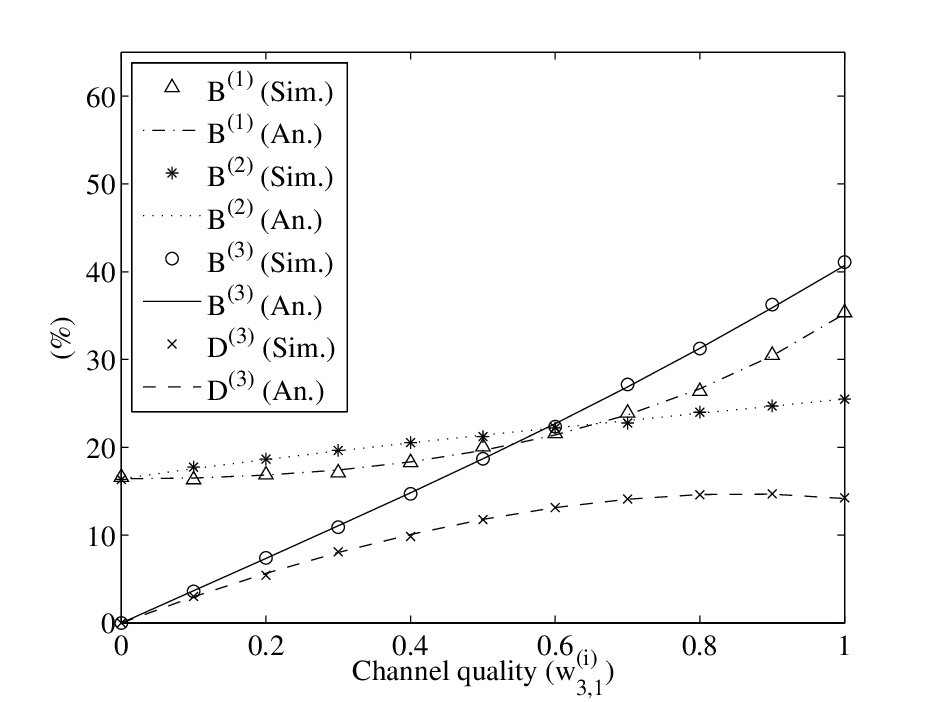}\label{fig:incr_w31_K=2}}
\subfigure[$K=3$]{\includegraphics[width=0.49\columnwidth]{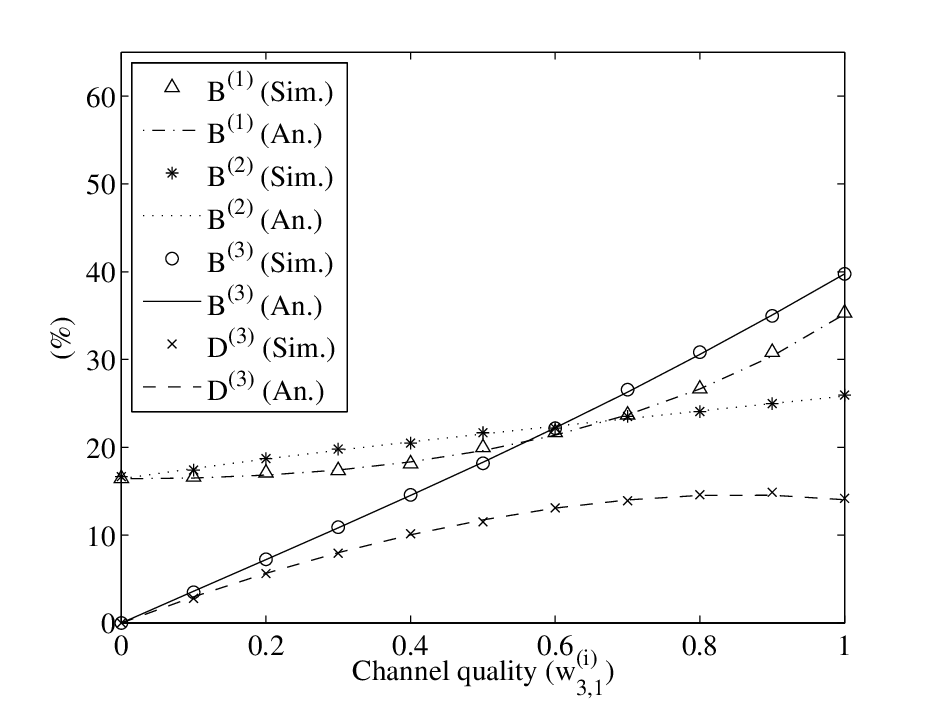}\label{fig:incr_w31_K=3}}
\caption{Illustration of the relationship of the blocking probability for group 3 UEs, $B^{(3)}$, the collision probability for group 3 UEs, $D^{(3)}$, the blocking probability for group 1 UEs, $B^{(1)}$, and the blocking probability for group 2 UEs, $B^{(2)}$, as a function of the channel quality between group 3 UEs and BS 1, $w_{3,1}^{(i)}$ and four all possible shared channels, i.e. (a) $K=0$, (b) $K=1$, (c) $K=2$, (d) $K=3$. We observe that the primary factor in the degradation of performance for group 3 UEs in this particular scenario is $B^{(3)}$ as compared to $D^{(3)}$ and this difference increases as $w_{3,1}^{(i)}$ improves. Once again, this figure shows good agreement between the results from the analytical model and from simulation. }
\label{fig:incr_w31}
\end{figure*}

In Fig.~\ref{fig:incr_w31} we examine $B^{(3)}$, $D^{(3)}$, $B^{(1)}$, and $B^{(2)}$, as a function of $w_{3,1}^{(i)}$ for all possible shared channels, i.e. $K=\{0,1,2,3\}$. The scenario used in this result is identical to the one used previously. By comparing an increasing value of $K$ in Fig.~\ref{fig:incr_w31_K=0}--Fig.~\ref{fig:incr_w31_K=3} we observe the impact of an increasing number of shared channels on the considered performance benchmarks. The first interesting observation is that irrespective of the value of $K$ the blocking probability for group 1 UEs, $B^{(1)}$, is relatively constant. This means that the quality of service requirements for UEs not involved in load balancing will be met, even with load balancing enabled. Second, as $K$ increases, blocking probability for group 3 UEs, $B^{(3)}$, significantly decreases, which proves the effectiveness of load balancing in this context. Furthermore, the collision probability for UEs in group 3, $D^{(3)}$, also reduces because with more shared channels there are fewer unconnected UEs to request connections. Note, however, that the difference in collision probability for an increasing $K$ is not as significant as observed for the blocking probability because increasing the number of shared channels has a minimal effect on the performance of the random access phase. Finally, increasing $K$ only slightly increases blocking probability $B^{(2)}$ because these UEs have priority in connecting to any of cell 2's free channels.

Focusing on Fig.~\ref{fig:incr_w31_K=3} only, where load balancing is enabled, we note that as $w_{3,1}$ increases, all curves experience an increase. This can be explained as follows: with an increase in $w_{3,1}^{(i)}$, more group 3 UEs are able to generate successful connections to BS 1 resulting in an increase in the contention for sub slots during admission control, and hence an increase in $D^{(3)}$. Also, there is an accompanied increase in $B^{(3)}$ because as more UEs generate successful connections, an increasing number of UEs contend for free channels on both cell 1 (where load balancing does not occur) and cell 2 (where load balancing occurs). Consequently, this also results in an increase in $B^{(1)}$ and $B^{(2)}$. Although these trends are obvious, the exact degradation in UE experience for each group is not. For example, in this specific scenario, Fig.~\ref{fig:incr_w31} illustrates that $B^{(3)}$ is always the primary factor in the degradation of the group 3 UE experience as compared to $D^{(3)}$. This knowledge is significant as the network operator can determine whether an increase in $K$, or an increase $L_3$ will be more beneficial to group 3 UEs. Observe that Fig.~\ref{fig:chan_util_w31} and Fig.~\ref{fig:incr_w31} show an extremely good match between the analytical result and simulation.

\subsection{Impact of Random Access Phase on Load Balancing Process}
\label{ses:varying_p}

\begin{figure*}
\centering
\subfigure[]{\includegraphics[width=0.49\columnwidth]{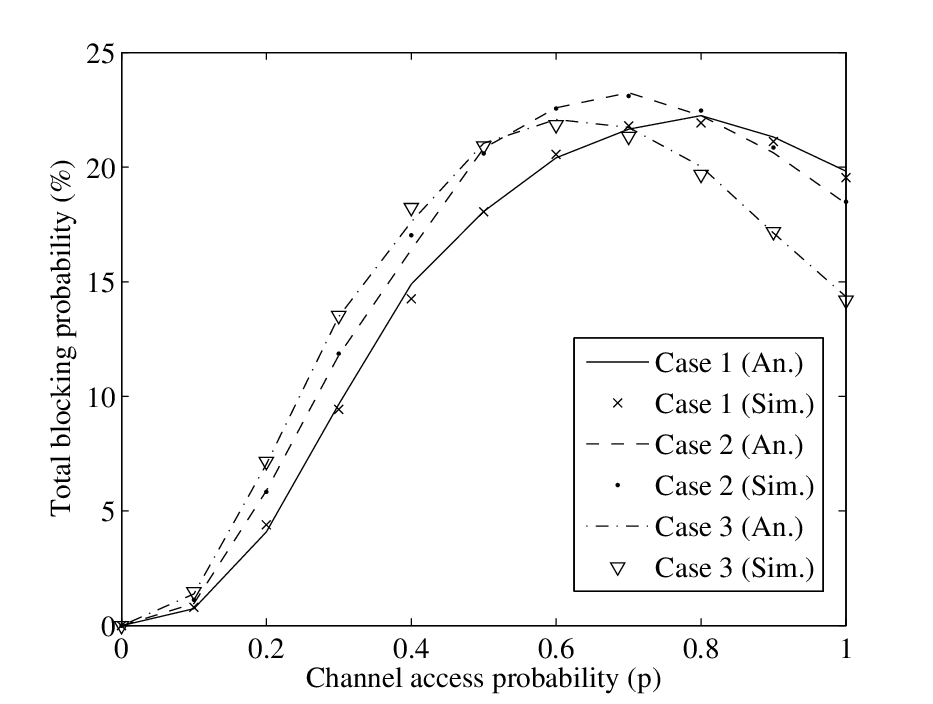}\label{fig:blocking_tot_p}}
\subfigure[]{\includegraphics[width=0.49\columnwidth]{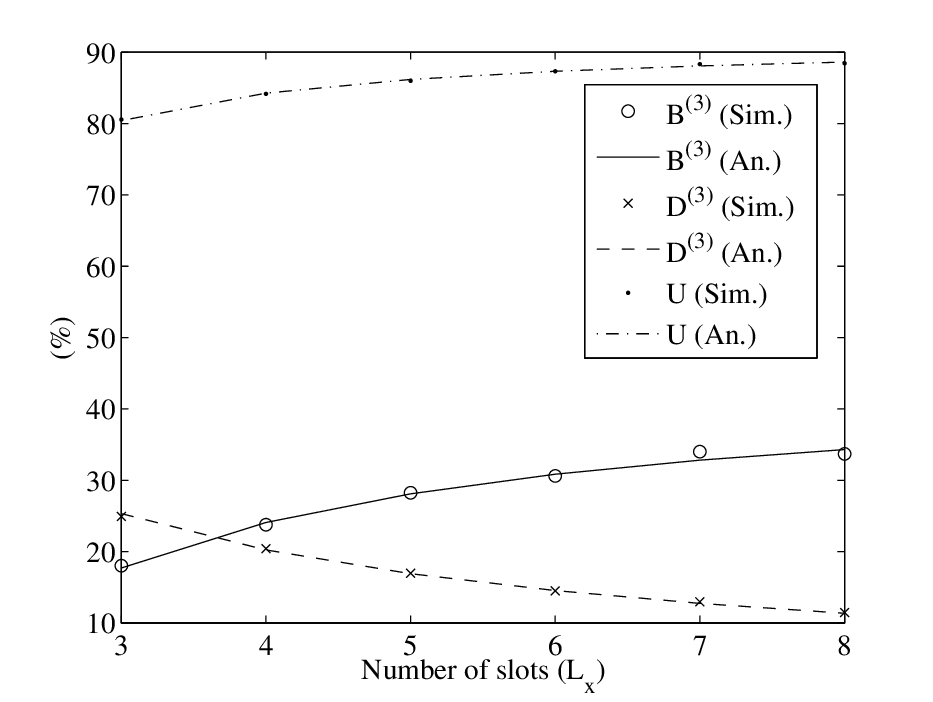}\label{fig:varying_L}}
\caption{Impact of random access phase on load balancing process: (a) total network-wide blocking probability as a function of access probability $p$; (b) impact of random access phase length $L_x$ on the performance metrics of the considered system. We observe that the blocking probability is not a monotonous function of $p$ and there is an extremum beyond which blocking starts to drop-off. As a result, the network metrics can be easily adapted by the network operator by dynamically selecting number of random access slots.}
\label{fig:access_effect}
\end{figure*}

In this section we present results on the effect of random access phase on the performance of load balancing. The results are presented in Fig.~\ref{fig:access_effect}. All network parameters are set identically to the network considered in Section~\ref{sec:cqi}, except for $L_x=3$, where $x\in\{1,2\}$.

We begin by investigating the impact of different UE distributions on the performance of load balancing. We perform three experiments and denote each experiment as a specific case. In the first case we set the number of UEs, such that more UEs are distributed in groups 1 and 2, than in group 3, i.e. $N_1=N_2=6$, $N_3=4$. In the second case we set the number of UEs equal in each group, i.e. $N_1=N_2=N_3=6$. And finally, in the third case we set the number of UEs in group 3 larger than in the other two groups, i.e. $N_1=N_2=6$, $N_3=8$. The metric that is studied in the three cases described above is the total network-wide blocking probability, calculated as $\frac{1}{3}\sum_{i=1}^{3}B^{(i)}$, as a function of channel access probability $p_i=p$ for $i\in\{1,2,3\}$. This metric is used in order to give a simple overall indication of the blocking suffered by UEs in all groups. Results are presented in Fig.~\ref{fig:blocking_tot_p}.

The most interesting observation from Fig.~\ref{fig:blocking_tot_p} is that with an increase in the number of UEs in the TTR, the total blocking probability becomes smaller for moderate values of $p$. Surprisingly, the blocking probability starts to drop sharply as the value of $p$ continues to increase. This phenomenon occurs because as $p$ increases, so do collisions on the random access channel, which in-turn limits the blocking probability because fewer UEs successfully access available channels. The result is easier to understand when one observes that the blocking probability is the probability of not finding a free data channel for a connection that has successfully connected to the BS via a control channel. It has to be kept in mind that for each case presented in Fig.~\ref{fig:blocking_tot_p} the length of the random access phase remains the same. What is important to note is that for moderate values of $p$, the difference between blocking probabilities for each case is small, i.e. less than 5\% (please compare values of blocking probability for each case in the range of $p\in(0,0.6)$). However, as $p$ becomes very large, the curves with a higher number of group 3 UEs drop off faster because they experience a substantial increase in the number of collisions. Therefore, a certain value of $p$ can maximize the channel utilization achieved through load balancing and also maintain the blocking probability at approximately the same level (given negligible changes in UE distribution).

We now move our focus to the impact of random access phase length $L_x$ on the performance of load balancing. The results are presented in Fig.~\ref{fig:varying_L}. The set of parameters remain the same as in the earlier experiment in this section, however $p_x=p=0.4$. As an example, three network metrics are evaluated as a function of number of slots in the random access phase, $L_x$: (i) total channel utilization in both cells, $U$, (ii) collision probability at group 3, $D^{(3)}$, and (iii) blocking probability at group 3, $B^{(3)}$. For simplicity, the number of slots is set equal among each group of UEs.

Obviously, as the number of random access slots increase the collision probability decreases for group 3 UEs, and the overall channel utilization increases. However, as the collision probability decreases the blocking probability, within the same group of UEs, becomes larger. This is in line with the results presented in Fig.~\ref{fig:blocking_tot_p}. Recall, that as more UEs gain access to the BS, the probability that channels become unavailable increases. The results shown in Fig.~\ref{fig:varying_L} further demonstrate the fundamental tradeoff between the delay caused by random access and overall network utilization. With an increase in traffic intensity $p$, we expect that the graphs shown in Fig.~\ref{fig:varying_L} to shift upwards proportional to the increase in $p$ because we expect that a higher traffic intensity would result in more collisions and blocking for users in group 3. We demonstrate that the network operator has a powerful tool, i.e. random access phase length, through which network metrics can be easily regulated. It is obvious that the operator has no control over the channel access probability, $p$, of individual UEs. However, the operator is able to set a higher value of $L_x$ to the ASC of interest in order to maintain an expected access delay for each UE against a required blocking probability. 

\subsection{Impact of Varying Shared Channel Pool $K$ on Load balancing Efficiency of DR$_K$}
\label{sec:varying_k}

We consider a macrocell scenario in which the distribution of UEs in groups 1, 2 and 3 follow the relationship given by $N_1=N_2<N_3$ and $L_x=N_x$, where $x\in\{1,2,3\}$. Let $N_1=N_2=L_1=L_2=25$ and $N_3=L_3=40$. We consider a symmetric system where each cell has $10$ channels, i.e. $M_1=M_2=10$, and the distances between each group of UEs and their respective serving BSs are identical, i.e. $d_{1,1}=d_{3,1}=d_{3,2}=d_{2,2}=220$\,m. Once again, we assume an average channel capacity of $R=250$\,kbps per channel with an average packet size $r_p=1$\,kB. A frame duration duration of $\tau=1$\,ms is used. We assume a simplified path loss model with identical parameters as in Section~\ref{sec:cqi} except with $\delta=3$, which is more appropriate for outdoor channel conditions.
\begin{figure*}
\centering
\subfigure[]{\includegraphics[width=0.49\columnwidth]{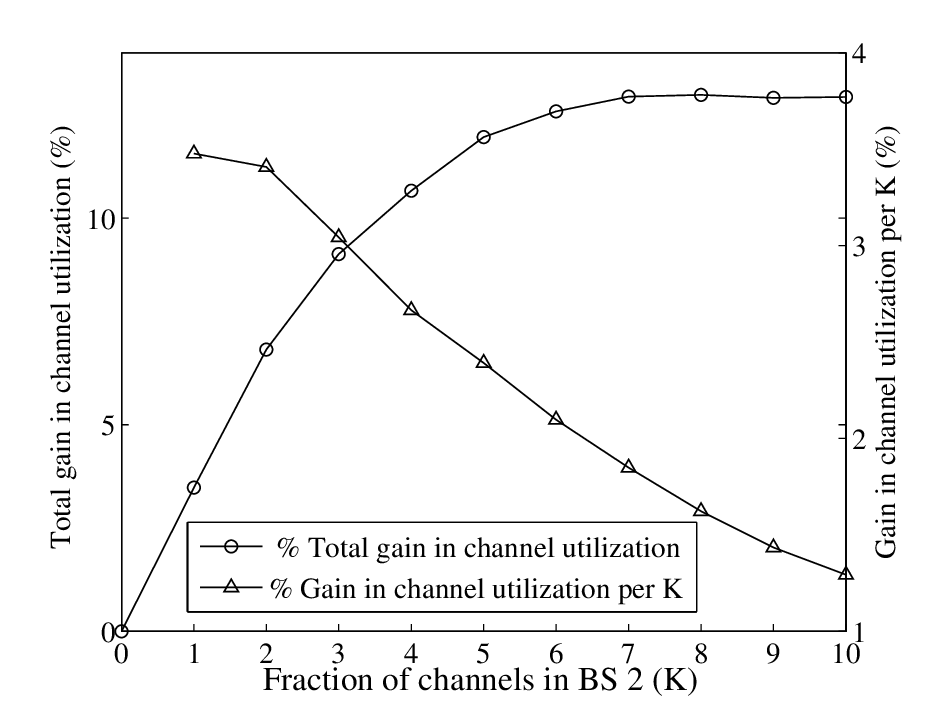}\label{fig:ch_util_change}}
\subfigure[]{\includegraphics[width=0.49\columnwidth]{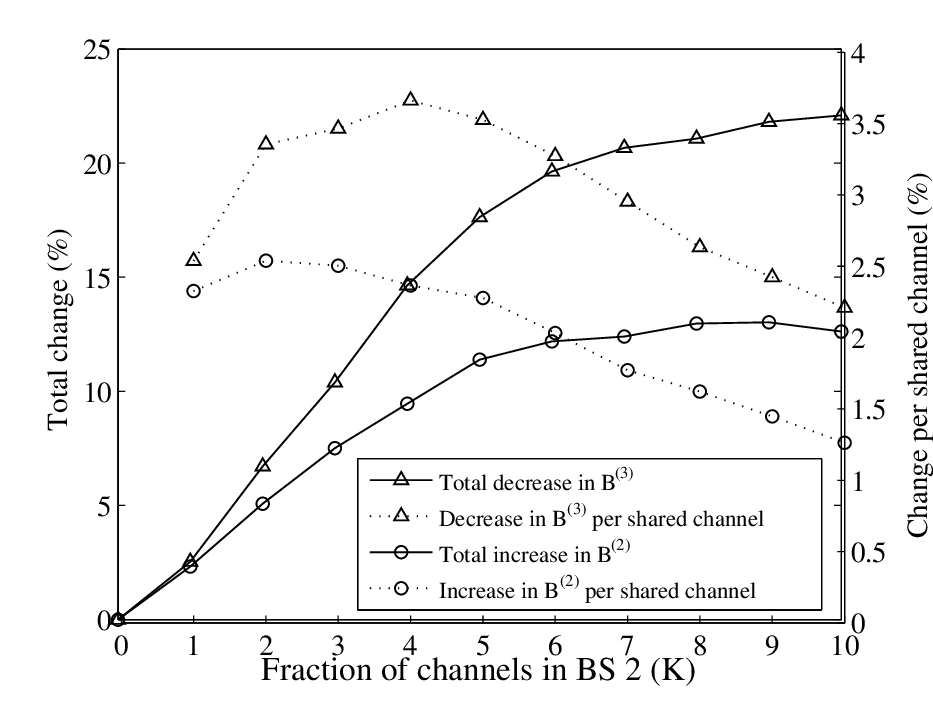}\label{fig:block_prob}}
\caption{Impact of $K$ on load balancing efficiency: (a) The percentage improvement of channel utilization (left vertical axis) and the percentage improvement of channel utilization per shared channel (right vertical axis) as a function of shared channels, $K$. As $K$ increases, there is an obvious improvement in channel utilization, however, there are decreasing gains experienced per additional $K$; (b) The total change in blocking probability, $B^{(x)}$, and blocking probability per shared channel, $B^{(x)}/K$, where $x \in \left\{ {2,3} \right\}$. In this scenario, we observe that the decrease in $B^{(3)}$ is always greater than the corresponding increase in $B^{(2)}$ for all $K$, suggesting that there is an overall improvement in the UE experience. Furthermore, it is seen that at $K=6$ we have the greatest difference between the decrease in $B^{(2)}/K$ and the corresponding increase in $B^{(2)}/K$ suggesting that this is the optimal number of shared channels to use in order to gain the best UE experience per $K$.}
\end{figure*}

In Fig.~\ref{fig:ch_util_change} we explore the system-wide improvement in channel utilization (represented as a percentage on the vertical axis) as a function of the number of shared channels $K$ (represented on the horizontal axis). Note that for the remainder of our study we fix traffic intensity for all groups to $p=0.2$ in order to determine the maximum gain in channel utilization for every value of $K$ at traffic intensities that are very near saturation. The line with circle markers represents the total percentage improvement experienced as a function of $K$, while the line with triangle markers represents the improvement experienced in channel utilization per shared channel. As $K$ increases, there is a decreasing improvement in channel utilization per additional channel, which indicates that the extra cost of sharing more channels for load balancing may outweigh the added benefit of serving a greater number of UEs.

Although there is an overall improvement in the system-wide channel utilization, the exact effect of the load-balancing scheme on the UE experience is unknown. Obviously, $B^{(3)}$ decreases with an increase in $K$ because group 3 has access to more channels. In contrast, $B^{(2)}$ increases because more UEs in group 3 access channels belonging to cell 2, which are of course also accessible to UEs in group 2. Although these general trends are obvious, the exact relationship between the amount of decrease in $B^{(3)}$ versus the amount of increase in $B^{(2)}$ is unknown. In Fig.~\ref{fig:block_prob} we examine this relationship in more detail, where the decrease in $B^{(3)}$ (solid line with triangle markers) is plotted with the consequent increase in $B^{(2)}$ (solid line with circles) as a percentage on the vertical axis with increasing $K$ on the horizontal axis. We observe that in this particular scenario the total decrease in $B^{(3)}$ is always more than the total increase in $B^{(2)}$, suggesting that the overall UE experience improves with the proposed load-balancing scheme. This reaffirms the increase in channel utilization with an increase in $K$ for DR$_K$, which is previously observed in Fig.~\ref{fig:ch_util_change}.

The total improvement in overall UE blocking probability demonstrates the effectiveness of the load balancing scheme. However, from a network operator viewpoint, knowledge of the changes in UE experience per additional shared channel is also very important. Fig.~\ref{fig:block_prob} examines the effect of an increase in $K$ on both the decrease in $B^{(3)}/K$ (dashed line with triangles), and the consequent increase in $B^{(2)}/K$ (dashed line with circles). We observe, that for this particular scenario, a decrease in $B^{(3)}/K$ is always more than the increase in $B^{(2)}/K$, which suggests that the UEs in group 3 experience more of an improvement in performance than the performance degradation experienced by UEs in group 2 per additional shared channel. This allows direct evaluation of the effectiveness of the load-balancing scheme on the overall UE experience per additional shared channel. In Fig.~\ref{fig:block_prob}, we note that $B^{(3)}/K$ reaches a maximum at an intermediate value of $K$, i.e. $K=3$, because the system reaches a balance between the number of UEs requesting connections and those that are already connected through load balancing. Our model allows for the direct observation of this system state because of the combined modeling of a finite number of UEs together with a detailed call admission process. With the use of Fig.~\ref{fig:block_prob}, we are able to determine the best $K$ to improve the overall UE experience on a per shared channel basis, and then find the corresponding improvement in the overall channel utilization using Fig.~\ref{fig:ch_util_change}. For this particular scenario, the maximum difference between the increase in $B^{(3)}/K$ and decrease in $B^{(2)}/K$ occurs at $K=6$ and corresponds to an overall improvement in channel utilization of 12.6\%.

In summary, we can construct the following optimization function. Given fundamental descriptors of the network considered in Section~\ref{sec:numerical_analysis}, i.e. $N_i$, $M_i$, $L_x$, $p_i$, $w_{x,y}^{(i)}$, $r_p$, and $\tau$, the network operator should find
\begin{equation}
\arg \max_K U/K \text{~subject to~} \forall i B^{(i)}\leq m^{(i)}, D^{(i)}\leq o^{(i)},
\label{eq:optimization}
\end{equation}
where $m^{(i)}$ and $o^{(i)}$ are the required maximum blocking and collision probabilities, respectively, for group $i$. The developed analytical model provided in Section~\ref{sec:numerical_analysis} allows solving the optimization function (\ref{eq:optimization}), since each metric, $U$, $B^{(i)}$ and $D^{(i)}$ is given in closed form. The optimization formula allows obtaining the value of $K$ in order to obtain the maximum utilization per shared channel, such that all considered quality of service metrics required by the operator, $m_i$ and $o_i$, are met. Note that finding the optimal solution to (\ref{eq:optimization}) is beyond the scope of this paper. On the other hand, note that since the complexity of the optimization problem (\ref{eq:optimization}) increases linearly with $K$, it can be solved efficiently via, e.g., exhaustive search, provided the existence of quick computation of $\pi_{a,b,c,d}$. Then, the complexity of calculating $\pi_{a,b,c,d}$ increases nearly-exponentially with increasing $M_i$. Note however that calculating $\pi_{a,b,c,d}$ even for large values of state space (e.g., for $K=3$, $M_1=M_2=12$ and $N_x=24$ size of  $\pi_{a,b,c,d}$ is 16,744 elements), can be handled efficiently with any standard personal computer with optimized implementation of the transition probability matrix developed in this paper.

\subsection{Load balancing with OSA Results}
\label{sec:osa_results}

We consider an energy detection technique to detect the presence of cell 2 users and an AWGN channel for which $\mu_f$ and $\mu_m$ is given as~\cite[Sec. III]{digham_icc_2003}. The sensing bandwidth is 200\,kHz, with cell 2 users detected SNR of --5\,dB and the detection threshold of --109.4\,dBm is set to the noise floor.

\subsubsection{Impact of Varying $M_2$ on Throughput for Group 3 Users $Z$}
\label{sec:impacts_M2}

To illustrate an application of the model we consider the following parameters: the number of channels on cell 1 $M_1=7$, the number of subscribers in the region of overlap $N_3=40$, connection request probability $p=0.1$, random access phase length $L=N_3/2=20$. Furthermore, we consider an average packet size of $r_p=1$\,kB, channel throughput $R=250$\,kbps and a transmission time of $\tau-\tau_s=9$\,ms, where $\tau_s=1$\,ms. This results in an inverse of the packet length $q=0.2813$, false alarm probability $\mu_f=0.1398$ and mis-detection probability $\mu_m=0.0861$.

\begin{figure*}
\centering
\subfigure[]{\includegraphics[width=0.49\columnwidth]{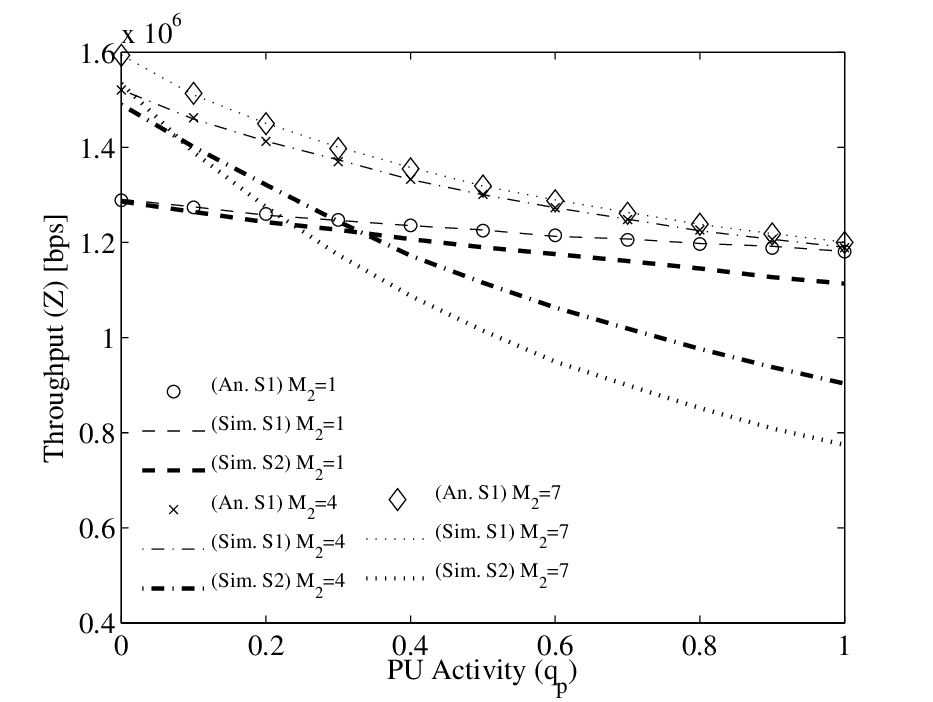}\label{fig:chutil_M2_qp}}
\subfigure[]{\includegraphics[width=0.49\columnwidth]{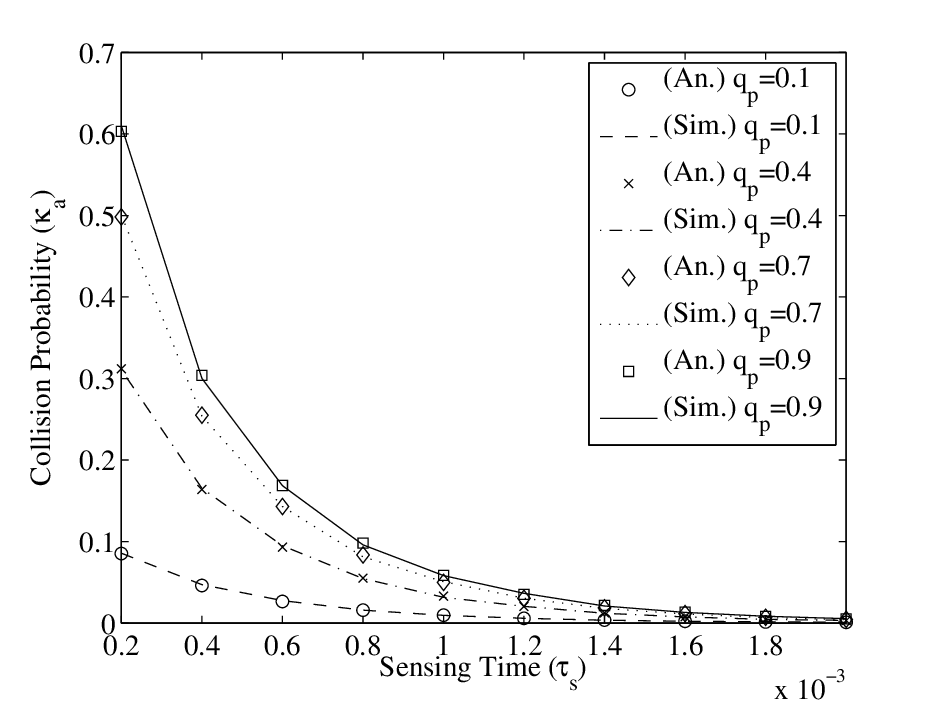}\label{fig:coll_qp_Ts}}
\caption{Performance of OSA load balancing system: (a) Group 3 users throughput, $Z$, as a function of cell 2 user activity, $q_p$, for varying number of channels accessible to group 3 users on cell 2, $M_2$. The labels (An. S1) and (Sim. S1) represent the analytical and simulation results, respectively, for the DR$_K$ load balancing scheme. (Sim. S2) represents the results from a random cell 1/cell 2 channel selection algorithm as analyzed in~\cite{almahdi_commlet_2009}. We observe a decreasing marginal gain in throughput as $M_2$ increases with significant gains experienced at relatively low values of $q_p$. Note the perfect match between analysis and simulations of system S1; (b) Average cell 1/cell 2 users collision probability, $\kappa_a$, as a function of the sensing time, $\tau_s$, for varying levels of cell 2 users activity, $q_p$. An increase in $\tau_s$ to reduce $\kappa_a$ is more effective at high levels of $q_p$ where the system is highly sensitive to changes in the probability of mis-detection, $\mu_m$. Note that just as in Fig.~\ref{fig:chutil_M2_qp} there is a perfect match between analysis and simulations.}
\end{figure*}

Results associated with the above parameters are presented in Fig.~\ref{fig:chutil_M2_qp}. For comparison we also plot the throughput using a random cell 1/cell 2 channel selection algorithm for new group 3 user connections, which was analyzed in~\cite{almahdi_commlet_2009}. As expected, in both cases group 3 user throughput decreases with an increase in cell 2 user activity. This decrease in throughput is attributed to more group 3 user preemptions and blocked connections. Furthermore, group 3 user throughput increases with an increase in $M_2$. More interestingly, as $M_2$ increases the OSA-enabled cellular network experiences less marginal gain in throughput for each additional group 3 user accessible channel on cell 2. Moreover, the greatest gain in throughput for additional $M_2$ channels is seen at low values of $q_p$, e.g. $q_p<0.2$. With such knowledge operators have insight on the marginal gains in throughput with additional $M_2$ and the specific role $q_p$ plays in limiting these gains. The solution in the present paper based on first assigning channels from cell 1 results in a significant improvements over the random assignment approach. For example, for $M_2=7$ at $q_p=0.7$ our algorithm provides more than one and a half times improvement.

\subsubsection{Impact of Sensing Time $\tau_s$ on cell 1/cell 2 users Collision Probability $\kappa_a$}
\label{sec:impacts_coll}

The impact of $\tau_s$ on the collision probability of cell 1/cell 2 users is shown in Fig.~\ref{fig:coll_qp_Ts}. All parameters remain the same as in Section~\ref{sec:impacts_M2}, except that the sensing threshold equals $-109.7$\,dBm, $r_p=800$\,kB and $M_1=1$, $M_2=6$ for the purpose of better illustrating the impact of $\tau_s$. More collisions with an increase in $q_p$ are experienced because more mis-detections of cell 2 occur. Furthermore, an increase in $\tau_s$ results in an accompanying decrease in $\mu_m$ thus reducing $\kappa_a$ due to lower levels of perceived cell 2 user activity. In addition, increasing $\tau_s$ to reduce the number cell 1/cell 2 users collisions is more effective at high $q_p$. At high $q_p$ the system is highly sensitive to $\mu_m$, i.e. a small reduction in $\mu_m$ results in large improvements. Therefore, network operators need to account for relative levels of $q_p$ to determine if increasing $\tau_s$ will result in a considerable improvement for cell 2 users. At low $q_p$, network operators can determine the optimal $\tau_s$ to maximize $Z$ while measuring the exact improvement in $\kappa_a$.

\subsection{Extension of the Model to a Multi-Cell Scenario}
\label{sec:multicell}

Due to the complexity of deriving performance metrics based on an analytical model for a multi-cell network, we present a numerical approximation and simulation results to provide insight into the behavior of such a system. We consider a general, non-OSA, scenario in which a central cell overlaps with $X$ neighboring cells. UEs are divided into groups, in the same fashion as in the system model presented earlier. Specifically, there are three groups: (i) UEs in the central cell (referred to as group 1), (ii) UEs in the neighboring cell (referred to as group 2) and (iii) UEs in the overlap region between the central and neighboring cells (referred to as group 3). It is assumed that UEs in the central cell and in each of the overlapping regions are registered to the central cell. To avoid ambiguity we assume that there are no overlaps between cells neighboring the center cell, so that the UEs in the overlap region observe signals only from two cells. This assumption makes evaluation of the multi-cell extension relatively easier.

Each of the groups considered consists of $N$ UEs, while every cell has $M$ available channels. Therefore, a particular UE belonging to any group in the overlap region has access to $2M$ possibly unoccupied channels, i.e. $M$ from the central cell and an additional $M$ from the neighboring cell, while the UEs from the remaining groups have access to only $M$ channels. Each channel is assumed to be error free (in other words channel conditions are not considered). Note that the decision on which users are granted connections to the central cell in the presence of load balancing is left solely to the radio network controller (as in the case of base two-cell case, refer to Section~\ref{sec:load_balancing_Process} for details). In the case of our analysis we apply a simple decision strategy where users are selected randomly from each cell when load balancing needs to occur. In practical scenarios the radio network controller can apply user selection strategies based on, e.g., (i) user/cell priority, (ii) fairness, and/or (iii) channel quality. We present the results on two essential metrics: (i) the overall channel utilization, and (ii) the overall blocking probability.

We utilize a simple numerical approximation for overall channel utilization. We represent the two-cell system, as a special case of a multi-cell system, which has one central cell and $X$ neighboring cells. For simplicity, in this particular case, we assume that only group 3 UEs are present in the multi-cell system. Furthermore, cell one of the two-cell system represents the central cell of the multi-cell system, while cell two represents a linear combination of the $X$ neighboring cells in the multi-cell system. The number of UEs in the group residing in the region of overlap of the two-cell system is now $XN$, the number of channels in each cell is $\lfloor (XM)/2\rfloor$ and the number of assigned random access slots is $XL$. In this manner, we keep the ratio of the number of UEs per channel approximately the same for both the two-cell and multi-cell systems. We present the results for three network scenarios: (i) a large-scale network, with $M=15$ and $N=30$, (ii) a medium-scale network, with $M=8$, $N=16$ and (iii) a small-scale network with $M=3$, $N=6$. For each scenario we assume a channel access probability $p=0.4$, average channel capacity of $R=250$\,kbps per channel with an average packet size $r_p=0.833$\,kB and $\tau=8$\,ms slot length (the same parameters as in Section~\ref{sec:cqi}) which translates to a slot occupancy probability of $q=0.3$. Since the traffic generated by users in each cell neighboring the central cell is the same, we keep $K=M$ for every cell. Note the an unequal pool of shared channels would penalize some cells during the load balancing process and defeat the goal of maintaining an equal level of throughput for each cell.

\begin{figure*}
\centering
\subfigure[]{\includegraphics[width=0.49\columnwidth]{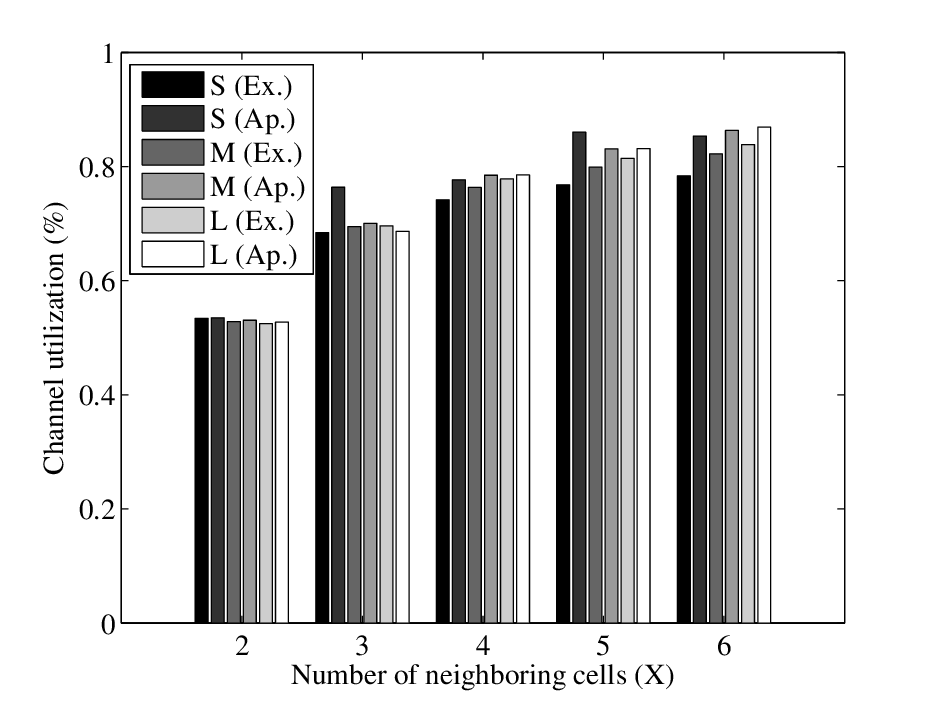}\label{fig:ch_util_approx}}
\subfigure[]{\includegraphics[width=0.49\columnwidth]{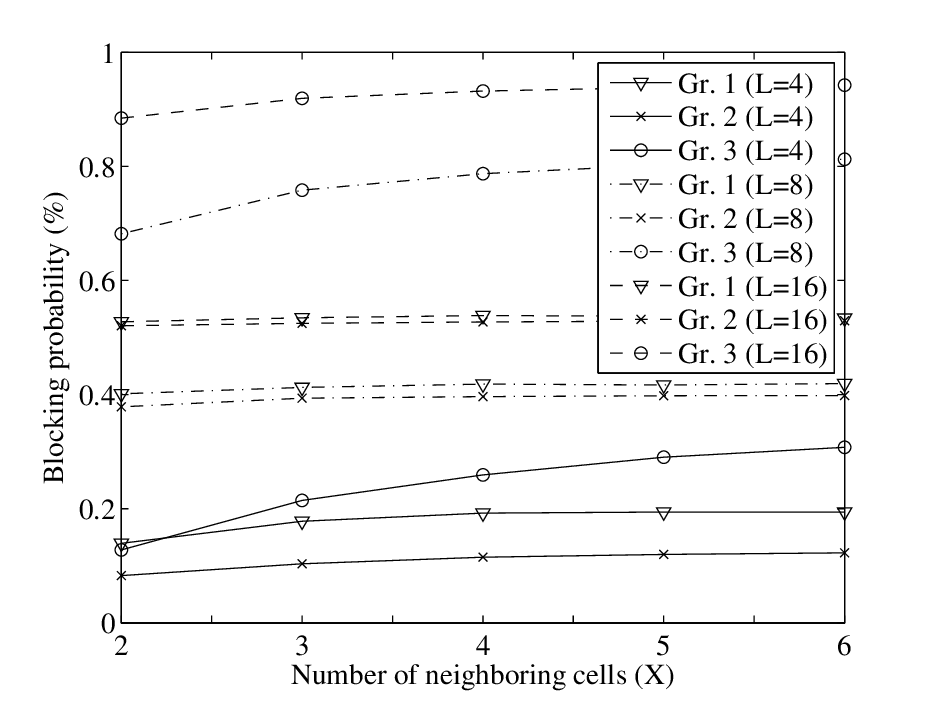}\label{fig:blocking_X}}
\caption{Results for the multi-cell system: (a) channel utilization (Ex.: exact; Ap.: approximation; S: small; M: medium; L: large) and (b) simulation results for blocking probability (Gr.: Group). The parameter setup and explanation of the naming convention is presented in Section~\ref{sec:multicell}. We also note that the channel utilization in the multi-cell system does not drastically differ as the number of UEs and channels change. Also, the blocking probability for each group increases with an increase in the number of neighboring cells. The blocking probability experiences the most increase for UEs in the overlap region (group 3).}
\label{fig:approx}
\end{figure*}

In Fig.~\ref{fig:ch_util_approx}, we observe that as the number of neighboring cells increases, so does the channel utilization. Interestingly, the channel utilization does not change considerably with an increase in UE population and channel pool. Observe that the channel utilization is well represented by our approximation, i.e. the difference between the exact model and the approximation does not exceed 10\% for all network setups. Note that the approximation is looser for the small-scale network, in Fig.~\ref{fig:ch_util_approx} for $X=\{3,5\}$, i.e. when the ratio of UEs to channels is not an integer. This is simply due to the rounding down of the number of the channels by the floor function.

In Fig.~\ref{fig:blocking_X} we present simulation results on the blocking probability as a function of the number of the neighboring cells $X$. We assume the same UE distribution as that of the medium-scale network, i.e. $M=8$, $N=16$, with $p=0.3$ with $r_p=0.625$\,kB, $R=250$\,kB, $\tau=8$\,ms (the same values as considered earlier), which translates to $q=0.4$. We observe the average blocking probability for each considered group in the scenario using three different values for the random access phase length $L=\{4, 8, 16\}$. The blocking probability for UEs in group 3 is more than that of the other groups with an increase in $X$. This is attributed to the increase in the number of UEs in the overlap region with an increase in the number of neighboring cells. The blocking probability for UEs in groups 1 and 2 stays relatively constant with changes in $X$ because UEs in group 1 have priority over UEs in group 3, while the UEs in group 2 do compete for cell resources with the UEs in the overlap region. Also, the relative difference between blocking for UEs in groups 1 and 2 for each value of $L$ is small because the ratio of accessible channels per UE remains the same regardless of $X$. As the number of random access slots increase, so does the blocking probability, which is fully consistent with the earlier observation expressed analytically and presented in Fig.~\ref{fig:varying_L}. This supports a conclusion that the analysis of the two-cell system is highly relevant to the behavior of the multi-cell network.

\section{Conclusions}
\label{sec:conclusions}

We have presented a new analytical model to assess the performance of the load balancing process in a two-cell system, later extended via approximations to a multi-cell system. The model differs in many respects from previous work on load balancing analysis in several important ways. First, it aids in determining the impact of channel quality to improve the accuracy of reported load balancing efficiency. Second, it demonstrates that the random access phase length can be used as an important network regulatory tool to control both system-wide and user experience performance metrics. Lastly, it facilitates the exploration of the effects of varying shared channel access on system-wide performance and user experience. In addition we have presented a new application of our model which allows for the consideration of new load balancing techniques. In specific, we have considered Opportunistic Spectrum Access (OSA) as an additional feature in the load balancing process. We have shown that OSA brings a large benefit to the cellular network, especially in the absence of central coordination. We have presented a variety of results derived from this framework, and in particular explored the tradeoffs in terms of channel utilization, blocking probability, and collision probability when traffic is transferred from a highly congested cell to a less-loaded neighboring cell.

\appendices

\section{Derivation of Transition Probabilities for the General Model}
\label{sec:appendix}

Before presenting the general solution, we introduce supporting functions that simplify the description of the transition probabilities. First, because the system is composed of three groups, where groups 1 and 2 have the same level of priority, we define a function that governs the transition probabilities for these groups, i.e.
\begin{equation}
\!\!\!{_{(M_{i},N_{j})}\alpha_{i_{t-1},j_{t-1}}^{(i_{t},z)}}\!\!\triangleq\!\!
\begin{cases}
\begin{split}&\sum_{k=0}^{i_{t}}T_{i_{t-1}}^{(k+i_{t-1}-i_{t})}\\&\times S_{i_{t-1}}^{(i)},\end{split} &\!\!\!\!\!\!\!\!\!\!\!\! \begin{split}i_{t}+j_{t} \leq M_{i},\\ i_{t}<i_{t-1};\end{split}\\
\begin{split}&\sum_{k=0}^{i_{t}}T_{i_{t-1}}^{(k)}\\&\times S_{i_{t-1}}^{(k+i_{t-1}-i_{t})},\end{split} &\!\!\!\!\!\!\!\!\!\!\!\!\begin{split}i_{t}+j_{t}<M_{i},\\ i_{t}\geq i_{t-1}~\text{or}\\ i_{t}+j_{t}=M_{i},\\ i_{t}\geq i_{t-1},\\j_{t}>j_{t-1};\end{split}\\
\begin{split}&\sum_{k=0}^{i_{t}}T_{i_{t-1}}^{(k)}\\&\times S_{i_{t-1}}^{(k+i_{t}-i_{t-1})}\\&+I_{j_{t-1}-z+i_{t}}^{(M_{i})}T_{i_{t-1}}^{k}\\&\times \sum_{n=i_{t}+1}^{N_{j}}S_{i_{t-1}}^{(k+n-i_{t-1})},\end{split}&\!\!\!\!\!\!\!\!\!\!\!\! 
\begin{split}i_{t}+j_{t} = M_{i},\\ i_{t}\geq i_{t-1},\\j_{t} \leq j_{t-1},\end{split}
\end{cases}
\label{eq:alpha}
\end{equation}
where $I_{i}^{(j)}=1$ when $i \geq j$ and $I_{i}^{(j)}=0$, otherwise. Variables $i_{x}$ and $j_{x}$ are supporting parameters that will be replaced by respective variables of (\ref{eq:rabcd}), once we derive general formulas for transition probabilities. The ranges of $i_{x}$ and $j_{x}$ will be defined in the respective transition probabilities shown later. Note that (\ref{eq:alpha}) resembles (\ref{eq:rat-1at}), except for the introduction of the indicator function $I_{i}^{(j)}$. For the remaining groups, we define the following supporting functions which denote the possible termination probabilities for group 3
\begin{equation}
{^{(n)}}\theta_{b_{t-1},c_{t-1}}^{(b_{t},c_{t},k,l)}\triangleq
\begin{cases}
T_{b_{t-1}}^{(k)}T_{c_{t-1}}^{(l)}, & n=1,\\
T_{b_{t-1}}^{(k+b_{t-1}-b_{t})}T_{c_{t-1}}^{(l+c_{t-1}-c_{t})}, & n=2,\\
T_{b_{t-1}}^{(k+b_{t-1}-b_{t})}T_{c_{t-1}}^{(l)}, & n=3,\\
T_{b_{t-1}}^{(k)}T_{c_{t-1}}^{(l+c_{t-1}-c_{t})}, & n=4.
\label{eq:theta}
\end{cases}
\end{equation}
Note that the termination probabilities for group 3 in (\ref{eq:theta}) are composed of individual termination function as given in (\ref{eq:tij}). One reason for this is that different channel qualities are experienced by UEs in the TTR, resulting in unequal termination probabilities, depending on whether these UEs are connected to BS 1 or BS 2. Lastly, we define
\begin{equation}
{^{(n)}}\xi_{b_{t-1},c_{t-1}}^{(b_{t},c_{t},k,l)}\triangleq
\begin{cases}
S_{b_{t-1}+c_{t-1}}^{(k+l+b_{t}+c_{t}-b_{t-1}-c_{t-1})}, & n=1,\\
S_{b_{t-1}+c_{t-1}}^{(k+l)}, & n=2,\\
S_{b_{t-1}+c_{t-1}}^{(k+l+c_{t}-c_{t-1})}, & n=3,\\
S_{b_{t-1}+c_{t-1}}^{(k+l+ b_{t}-b_{t-1})}, & n=4.
\label{eq:xi}
\end{cases}
\end{equation}
which denotes the possible arrangement probabilities for group 3. Note that the variables $k$, $l$ and $z$ in (\ref{eq:alpha}), (\ref{eq:theta}) and (\ref{eq:xi}) are the enumerators. Given the above, we can identify two major states of the system as follows: when all channels are occupied on both cells (called an edge state and having the same meaning as the last condition in (\ref{eq:rat-1at})); and the remaining states. We start by describing the edge state conditions.

\subsubsection{Edge state}
\label{sec:a+b=m}

Here we list the following sub-cases. For $a_{t}+b_{t}=M_{1}$, $c_{t}+d_{t}=M_{2}$, $b_{t} \geq b_{t-1}$, $c_{t} \geq c_{t-1}$ or $a_{t}+b_{t}=M_{1}$, $c_{t}+d_{t}<M_{2}$, $b_{t} \geq b_{t-1}$, $c_{t} \geq c_{t-1}$, $c_{t}=K$ we have
\begin{align}
\!\!r_{a_{t-1},b_{t-1},c_{t-1},d_{t-1}}^{(a_{t},b_{t},c_{t},d_{t})}&=\sum_{k=0}^{b_{t-1}}\sum_{l=0}^{c_{t-1}}\left({^{(1)}}\theta_{b_{t-1},c_{t-1}}^{(b_{t},c_{t},k,l)}{^{(1)}}\xi_{b_{t-1},c_{t-1}}^
{(b_{t},c_{t},k,l)}\nonumber\right.\\&\quad\left.+{^{(1)}}\theta_{b_{t-1},c_{t-1}}^{(b_{t},c_{t},k,l)}\sum_{r=l+1}^{N_{3}}{^{(1)}}\xi_{b_{t-1},c_{t-1}}^{(b_{t},c_{t},k,r)}\right)\nonumber\\&\quad\times{_{(M_{1},N_{1})}\alpha_{a_{t-1},b_{t-1}}^{(a_{t},k)}}\nonumber\\&\quad\times{_{(M_{2},N_{2})}\alpha_{d_{t-1},c_
{t-1}}^{(d_{t},l)}}.
\label{eq:13}
\end{align}
Equation (\ref{eq:13}) holds when the number of connections from group 3 UEs to both BS 1 and BS 2 increases. The first term in the brackets enumerates all the possible cases of terminations and generations in group 3 given a certain starting state. The second term in the brackets accounts for the edge case. This condition is similar in nature to the third case in (\ref{eq:rat-1at}). Lastly, the remaining ${^{(\cdot)}}\alpha_{(\cdot)}^{(\cdot)}$ terms account for the possible transitions in group 1 and 2. The indicator function used in the last condition of (\ref{eq:alpha}) is a function of the termination and connection enumerators in group 3. That is, depending on how many connections are admitted in BS 1 and BS 2 in a previous frame, a certain number of UEs from group 1 and 2 that request connections will not be admitted.

For $a_{t}+b_{t}=M_{1}$, $c_{t}+d_{t}=M_{2}$, $b_{t} <b_{t-1}, c_{t} < c_{t-1}$ or $a_{t}+b_{t}=M_{1}$, $c_{t}+d_{t}=M_{2}$, $b_{t}=b_{t-1}, c_{t}<c_{t-1}$ or $a_{t}+b_{t}=M_{1}$, $c_{t}+d_{t}=M_{2}$, $b_{t}<b_{t-1}, c_{t}=c_{t-1}$ or $a_{t}+b_{t}=M_{1}$, $c_{t}+d_{t}<M_{2}$, $b_{t}<b_{t-1}, c_{t}=c_{t-1}=K$ we have
\begin{align}
r_{a_{t-1},b_{t-1},c_{t-1},d_{t-1}}^{(a_{t},b_{t},c_{t},d_{t})}&=\sum_{k=0}^{b_{t}}\sum_{l=0}^{c_{t}}\left({^{(2)}}\theta_{b_{t-1},c_{t-1}}^{(b_{t},c_{t},k,l)}{^{(2)}}\xi_{b_{t-1},c_{t-1}}^{(b_
{t},c_{t},k,l)}\nonumber\right.\\&\quad+{^{(2)}}\theta_{b_{t-1},c_{t-1}}^{(b_{t},c_{t},k,l)}\sum_{r=l+1}^{N_{3}}\left.{^{(2)}}\xi_{b_{t-1},c_{t-1}}^{(b_{t},c_{t},k,r)}\right)\nonumber\\
&\quad\times{_{(M_{1},N_{1})}\alpha_{a_{t-1},b_{t-1}}^{(a_{t},k+b_{t-1}-b_{t})}}\nonumber\\
&\quad\times{_{(M_{2},N_{2})}\alpha_
{d_{t-1},c_{t-1}}^{(d_{t},l+c_{t-1}-c_{t})}}.
\label{eq:14}
\end{align}
In this case the number of connections from group 3 to both BS 1 and BS 2 decreases, or the number of connections in any one of the BSs remains the same, while the other decreases. The construction of the transition probability is the same as in (\ref{eq:13}), respectively replacing ${^{(1)}}\theta_{(\cdot)}^{(\cdot)}$ with ${^{(2)}}\theta_{(\cdot)}^{(\cdot)}$. Note that the definition of ${^{(2)}}\theta_{(\cdot)}^{(\cdot)}$ in (\ref{eq:theta}) defines the number of freed connections at BS 1 and BS 2 because of terminations of group 3 UEs. Since the number of connections have to be maintained at full cell occupancy for cell 1 and cell 2, the respective functions $_{(\cdot)}\alpha_{(\cdot)}^{(\cdot)}$ for BS 1 and BS 2, are used to compensate for the possible number of terminations at each cell due to group 3 UEs in order to maintain full system-wide occupancy, i.e. to remain at the edge state.

For $a_{t}+b_{t}=M_{1}$, $c_{t}+d_{t}=M_{2}$, $b_{t} <b_{t-1}, c_{t}>c_{t-1}$ or $a_{t}+b_{t}=M_{1}$, $c_{t}+d_{t}<M_{2}$, $b_{t} <b_{t-1}, c_{t}>c_{t-1}$, $c_{t}=K$
\begin{align}
r_{a_{t-1},b_{t-1},c_{t-1},d_{t-1}}^{(a_{t},b_{t},c_{t},d_{t})}&=\sum_{k=0}^{b_{t}}\sum_{l=0}^{c_{t-1}}\left({^{(3)}}\theta_{b_{t-1},c_{t-1}}^{(b_{t},c_{t},k,l)}{^{(3)}}\xi_{b_{t-1},c_{t-1}}^{(b_
{t},c_{t},k,l)}\nonumber\right.\\
&\quad+{^{(3)}}\theta_{b_{t-1},c_{t-1}}^{(b_{t},c_{t},k,l)}\sum_{r=l+1}^{N_{3}}\left.{^{(3)}}\xi_{b_{t-1},c_{t-1}}^{(b_{t},c_{t},k,r)}\right)\nonumber\\
&\quad\times{_{(M_{1},N_{1})}\alpha_{a_{t-1},b_{t-1}}^{(a_{t},k+b_{t-1}-b_{t})}}\nonumber\\
&\quad\times{_{(M_{2},N_{2})}\alpha_
{d_{t-1},c_{t-1}}^{(d_{t},l)}}.
\label{eq:15}
\end{align}
This case describes the situation where the number of connections from group 3 UEs to BS 1 strictly decreases, while those from group 3 UEs to BS 2 strictly increases. The transition probability represented in (\ref{eq:13}) can account for this by replacing ${^{(1)}}\theta_{(\cdot)}^{(\cdot)}$ with ${^{(3)}}\theta_{(\cdot)}^{(\cdot)}$. The definition of ${^{(3)}}\theta_{(\cdot)}^{(\cdot)}$ from (\ref{eq:theta}) describes the case when the number of terminations at BS 1 from group 3 UEs account for the decrease in the number of connections, while the number of terminations at BS 2 account only for any additional number of generations. The respective function $_{(\cdot)}\alpha_{(\cdot)}^{(\cdot)}$ for BS 1 and BS 2, again, must compensate for the changes in connections from group 3 to BS 1 and BS 2 in order to maintain full-cell occupancy.
Lastly, for $a_{t}+b_{t}=M_{1}$, $c_{t}+d_{t}=M_{2}$, $b_{t} >b_{t-1}, c_{t}< c_{t-1}$
\begin{align}
r_{a_{t-1},b_{t-1},c_{t-1},d_{t-1}}^{(a_{t},b_{t},c_{t},d_{t})}&=\sum_{k=0}^{b_{t-1}}\sum_{l=0}^{c_{t}}\left({^{(4)}}\theta_{b_{t-1},c_{t-1}}^{(b_{t},c_{t},k,l)}{^{(4)}}\xi_{b_{t-1},c_{t-1}}^{(b_
{t},c_{t},k,l)}\nonumber\right.\\&\quad+{^{(4)}}\theta_{b_{t-1},c_{t-1}}^{(b_{t},c_{t},k,l)}\sum_{r=l+1}^{N_{3}}\left.{^{(4)}}\xi_{b_{t-1},c_{t-1}}^{(b_{t},c_{t},k,r)}\right)\nonumber\\
&\quad\times{_{(M_{1},N_{1})}\alpha_{a_{t-1},b_{t-1}}^{(a_{t},k)}}\nonumber\\
&\quad\times{_{(M_{2},N_{2})}\alpha_{d_{t-1},c_
{t-1}}^{(d_{t},l+c_{t-1}-c_{t})}}.
\label{eq:16}
\end{align}
The above case is the opposite of that in (\ref{eq:15}). Here, the number of connections from group 3 to BS 1 strictly increases, while those from group 3 to BS 2 strictly decreases. Again, respective expressions for ${^{(1)}}\theta_{(\cdot)}^{(\cdot)}$ in (\ref{eq:13}) need to be replaced by ${^{(4)}}\theta_{(\cdot)}^{(\cdot)}$. The explanation for ${^{(4)}}\theta_{(\cdot)}^{(\cdot)}$ and $_{(\cdot)}\alpha_{(\cdot)}^{(\cdot)}$ given in (\ref{eq:theta}) is equivalent to the explanation for (\ref{eq:15}).

\subsubsection{Non-Edge state}

The second major group of cases refers to the situation in which the number of connections at BS 1 or BS 2 is less than or equal to the maximum capacity. This obviously involves more cases to consider than those explained in Section~\ref{sec:a+b=m}. We start by denoting conditions under which a transition from one state to another is not possible. That is, for $a_{t}+b_{t}< M_{1}$, $c_{t}+d_{t}\leq M_{2}$, $b_{t}>b_{t-1}$, $c_{t}>c_{t-1}$ or $a_{t}+b_{t}< M_{1}$, $c_{t}+d_{t}\leq M_{2}$, $b_{t}<b_{t-1}$, $c_{t}>c_{t-1}$
\begin{equation}
r_{a_{t-1},b_{t-1},c_{t-1},d_{t-1}}^{(a_{t},b_{t},c_{t},d_{t})}=0.
\end{equation}
For $a_{t}+b_{t}< M_{1}$, $c_{t}+d_{t}\leq M_{2}$, $b_{t}>b_{t-1}$, $c_{t}=c_{t-1}$
\begin{align}
r_{a_{t-1},b_{t-1},c_{t-1},d_{t-1}}^{(a_{t},b_{t},c_{t},d_{t})}&=\sum_{k=0}^{b_{t-1}}{^{(1)}}\theta_{b_{t-1},c_{t-1}}^{(b_{t},c_{t},k,0)}{^{(1)}}\xi_{b_{t-1},c_{t-1}}^{(b_{t},c_{t},k,0)}\nonumber\\
&\quad\times{_{(M_{1},N_{1})}\alpha_{a_{t-1},b_{t-1}}^{(a_{t},k)}}\nonumber\\&\quad\times{_{(M_{2},N_{2})}\alpha_{d_{t-1},c_{t-1}}^{(d_{t},0)}}.
\label{eq:18}
\end{align}
The above case is partially equivalent to (\ref{eq:13}) and considers the situation where the number of connections of group 3 UEs connected to BS 1 increases and those to BS 2 stay the same. Also, the number of new connections at BS 1 is less than its maximum capacity. Since this is not an edge case for the system, an additional third summation term is not needed as seen in (\ref{eq:13}). Note that this condition only contains one summation because the number of terminations from UEs connected to BS 2 cannot exceed the resultant connection state. This is because if they do exceed the desired number of terminations, UEs that generate connections to compensate for additional terminations will instead choose to connect to BS 1 (the BS they are registered to) thereby changing the resultant connection state.
Now, for $a_{t}+b_{t}=M_{1}$, $c_{t}+d_{t}< M_{2}$, $b_{t}\geq b_{t-1}$, $c_{t}\geq c_{t-1}$
\begin{align}
r_{a_{t-1},b_{t-1},c_{t-1},d_{t-1}}^{(a_{t},b_{t},c_{t},d_{t})}&=\sum_{k=0}^{b_{t-1}}\sum_{l=0}^{c_{t-1}}{^{(1)}}\theta_{b_{t-1},c_{t-1}}^{(b_{t},c_{t},k,l)}{^{(1)}}\xi_{b_{t-1},c_{t-1}}^{(b_{t},c_
{t},k,l)}\nonumber\\&\quad\times{_{(M_{1},N_{1})}\alpha_{a_{t-1},b_{t-1}}^{(a_{t},k)}}\nonumber\\
&\quad\times{_{(M_{2},N_{2})}\alpha_{d_{t-1},c_{t-1}}^{(d_{t},l)}}.
\label{eq:19}
\end{align}
This case is an extension of the case described in (\ref{eq:18}). However, full-cell occupancy now occurs at BS 1, i.e. all channels of BS 1 are occupied after the transition, and BS 2 operates at less than its maximum capacity. An additional summation is used as compared to (\ref{eq:18}) because full-cell occupancy on BS 1 allows for terminations to occur on BS 2 from group 3 UEs without changing their resultant connection number.
Now, for $a_{t}+b_{t}<M_{1}$, $c_{t}+d_{t}\leq M_{2}$, $c_{t}< c_{t-1}$, $b_{t}<b_{t-1}$, or $a_{t}+b_{t}=M_{1}$, $c_{t}+d_{t}< M_{2}$, $b_{t}=b_{t-1}$, $c_{t}<c_{t-1}$ or $a_{t}+b_{t}=M_{1}$, $c_{t}+d_{t}< M_{2}$, $b_{t}<b_{t-1}$, $c_{t}=c_{t-1}$
\begin{align}
r_{a_{t-1},b_{t-1},c_{t-1},d_{t-1}}^{(a_{t},b_{t},c_{t},d_{t})}&=\sum_{k=0}^{b_{t-1}}{^{(2)}}\theta_{b_{t-1},c_{t-1}}^{(b_{t},c_{t},k,0)}{^{(2)}}\xi_{b_{t-1},c_{t-1}}^{(b_{t},c_{t},k,0)}\nonumber\\
&\quad\times{_{(M_{1},N_{1})}\alpha_{a_{t-1},b_{t-1}}^{(a_{t},k+b_{t-1}-b_t)}}\nonumber\\
&\quad\times{_{(M_{2},N_{2})}\alpha_{d_{t-1},c_{t-1}}^{(d_{t},c_{t-1}-c_t)}}.
\label{eq:20}
\end{align}
The case described by (\ref{eq:20}) is a direct extension of (\ref{eq:14}). Similar to (\ref{eq:18}), terminations from group 3 UEs to BS 2 cannot be considered to achieve the resultant connection state.
Next, for $a_{t}+b_{t}=M_{1}$, $c_{t}+d_{t}< M_{2}$, $b_{t}<b_{t-1}$, $c_{t}< c_{t-1}$ or $a_{t}+b_{t}=M_{1}$, $c_{t}+d_{t}< M_{2}$, $b_{t}=b_{t-1}$, $c_{t}<c_{t-1}$ or $a_{t}+b_{t}=M_{1}$, $c_{t}+d_{t}< M_{2}$, $b_{t}<b_{t-1}$, $c_{t}=c_{t-1}$
\begin{align}
r_{a_{t-1},b_{t-1},c_{t-1},d_{t-1}}^{(a_{t},b_{t},c_{t},d_{t})}&=\sum_{k=0}^{b_{t}}\sum_{l=0}^{c_{t-1}}{^{(2)}}\theta_{b_{t-1},c_{t-1}}^{(b_{t},c_{t},k,l)}{^{(2)}}\xi_{b_{t-1},c_{t-1}}^{(b_{t},c_
{t},k,l)}\nonumber\\&\quad\times{_{(M_{1},N_{1})}\alpha_{a_{t-1},b_{t-1}}^{(a_{t},k+b_{t+1}-b_{t})}}\nonumber\\&\quad\times{_{(M_{2},N_{2})}\alpha_{d_{t-1},c_{t-1}}^{(d_{t},l+c_{t+1}-c_{t})}}.
\label{eq:21}
\end{align}
The above case is an extension of the transition probability described in (\ref{eq:20}). It considers the situation when full-cell occupancy occurs only at BS 1.
Next, for $a_{t}+b_{t}=M_{1}$, $c_{t}+d_{t}< M_{2}$, $b_{t}<b_{t-1}$, $c_{t}>c_{t-1}$
\begin{align}
r_{a_{t-1},b_{t-1},c_{t-1},d_{t-1}}^{(a_{t},b_{t},c_{t},d_{t})}&=\sum_{k=0}^{b_{t}}\sum_{l=0}^{c_{t-1}}{^{(3)}}\theta_{b_{t-1},c_{t-1}}^{(b_{t},c_{t},k,l)}{^{(3)}}\xi_{b_{t-1},c_{t-1}}^{(b_{t},c_
{t},k,l)}\nonumber\\&\quad\times{_{(M_{1},N_{1})}\alpha_{a_{t-1},b_{t-1}}^{(a_{t},k+b_{t+1}-b_{t})}}\nonumber\\&\quad\times{_{(M_{2},N_{2})}\alpha_{d_{t-1},c_{t-1}}^{(d_{t},l)}}.
\end{align}
The above case is an extension of the case described by (\ref{eq:15}). However, only full cell occupancy at BS 1 is considered.
Next, for $a_{t}+b_{t}<M_{1}$, $c_{t}+d_{t}\leq M_{2}$, $b_{t}>b_{t-1}$, $c_{t}<c_{t-1}$
\begin{align}
r_{a_{t-1},b_{t-1},c_{t-1},d_{t-1}}^{(a_{t},b_{t},c_{t},d_{t})}&=\sum_{k=0}^{b_{t}}{^{(4)}}\theta_{b_{t-1},c_{t-1}}^{(b_{t},c_{t},k,0)}{^{(4)}}\xi_{b_{t-1},c_{t-1}}^{(b_{t},c_{t},k,0)}\nonumber\\
&\quad\times{_{(M_{1},N_{1})}\alpha_{a_{t-1},b_{t-1}}^{(a_{t},k)}}\nonumber\\&\quad\times{_{(M_{2},N_{2})}\alpha_{d_{t-1},c_{t-1}}^{(d_{t},c_{t-1}-c_{t})}}.
\label{eq:23}
\end{align}
The above case is an extension of (\ref{eq:16}). In the case of (\ref{eq:23}) full-cell occupancy does not occur in any of the cells, therefore the respective summation terms from (\ref{eq:16}) accounting for the edge case are removed. Also, there is only one summation because as in (\ref{eq:18}) and (\ref{eq:20}) terminations from group 3 UEs to BS 2 cannot be considered to achieve the desired end state.
Lastly, for $a_{t}+b_{t}=M_{1}$, $c_{t}+d_{t}< M_{2}$, $b_{t}>b_{t-1}$, $c_{t}<c_{t-1}$
\begin{align}
r_{a_{t-1},b_{t-1},c_{t-1},d_{t-1}}^{(a_{t},b_{t},c_{t},d_{t})}&=\sum_{k=0}^{b_{t-1}}\sum_{l=0}^{c_{t}}{^{(4)}}\theta_{b_{t-1},c_{t-1}}^{(b_{t},c_{t},k,l)}{^{(4)}}\xi_{b_{t-1},c_{t-1}}^{(b_{t},c_
{t},k,l)}\nonumber\\&\quad\times{_{(M_{1},N_{1})}\alpha_{a_{t-1},b_{t-1}}^{(a_{t},k)}}\nonumber\\&\quad\times{_{(M_{2},N_{2})}\alpha_{d_{t-1},c_{t-1}}^{(d_{t},l+c_{t-1}-c_{t})}}.
\end{align}
The final case is an extension of the case described by (\ref{eq:23}). However, it considers the case when full cell occupancy occurs only at BS 1.

\section{Derivation of Transition Probabilities for the OSA Model}
\label{sec:appendix_osa}

First, we define a new indicator function as
\begin{equation}
^{(z)}I_{x}^{(y)}=
\begin{cases}
1,& \begin{split}\text{ if } x \ge y \text{ and } z = 1 \text{ or }\\ x = y \text{ and } z = 0,\end{split} \\
0,& \begin{split}\text{ if } x < y \text{ and } z = 1 \text{ or }\\ x \ne y \text{ and } z = 0.\end{split}
\end{cases}
\end{equation}

The transition probabilities are shown such that each case refers to specific changes in group 3 users connections on cells 1 and 2 from time slot $t-1$ to $t$. Thus, for $b_{t}<M_{1}$, $c_{t}+d_{t}\leq M_{2}$, $b_{t}\geq b_{t-1}$, $c_{t}=c_{t-1}$
\begin{align}
r_{b_{t-1},c_{t-1},d_{t-1}}^{(b_{t},c_{t},d_{t})}&=P_{c_t,d_t}^{(0)}\sum_{k=0}^{b_{t-1}}T_{b_{t-1}}^{(k)}T_{c_{t-1}}^{(0)}\nonumber\\&\quad\times S_{b_{t-1}+c_{t-1}}^{(k+b_{t}+c_{t}-b_{t-1}-c_{t-1})}.
\label{eq:5}
\end{align}
In (\ref{eq:5}) connections in cell 1 may increase, while the number of group 3 users connections on cell 2 does not change. The number of successful group 3 users connection generations must account for the possible number of terminations that can occur, as well as for the newly generated connections necessary to achieve the desired increase in the number of group 3 users connections on cell 1.

For $b_t<M_1 ,c_t+d_t \le M_2 ,b_t \ge b_{t-1} ,c_t>c_{t-1}$, as the number of group 3 users connections on cell 2 cannot increase without cell 1 being in the fully-connected state
\begin{equation}
r_{b_{t-1},c_{t-1},d_{t-1}}^{(b_{t},c_{t},d_{t})}=0.
\label{eq:6}
\end{equation}

For $b_t<M_1,c_t+d_t\le M_2, b_t<b_{t-1},c_t<c_{t-1}$ or $b_t<M_1,c_t+d_t\le M_2, b_t<b_{t-1},c_t=c_{t-1}$ or $b_t<M_1,c_t+d_t\le M_2, b_t = b_{t-1} ,c_t < c_{t-1}$
\begin{align}
\label{eq:7}
r_{b_{t-1},c_{t-1},d_{t-1}}^{(b_{t},c_{t},d_{t})}&=\sum_{i=0}^{i_m} P_{c_t,d_t }^{(i)} \sum_{k=0}^{b_{t-1}} T_{b_{t - 1}}^{(k + b_{t-1}-b_t)} T_{c_{t - 1}}^{(c_t-c_{t-1}-i)}\nonumber\\&\quad\times S_{b_{t-1}+c_{t-1}}^{(k)} {^{(1)}I_{c_{t-1}-c_{t}-i}^{(0)}},
\end{align}
where $i_m=\min\left(M_2-c_t,d_t\right)$. In (\ref{eq:7}) the number of group 3 users connections decreases on both cells or stays the same on one cell and decreases on the other. The termination probabilities are set to ensure a decrease occurs in the number of connections on both cells according to the state transition. Every possible number of terminations is considered and compensated for when regenerating connections on cell 1. Group 3 users connections on cell 2 compensate for every possible case of group 3 users preemption to ensure the exact number of terminations on cell 2. Note that every possible number of terminations on cell 2 is not accounted for since cell 1 is not in the fully-connected state.

For $b_t<M_1,c_t+d_t \le M_2,b_t > b_{t-1}, c_t < c_{t-1}$
\begin{align}
\label{eq:8}
r_{b_{t-1},c_{t-1},d_{t-1}}^{(b_{t},c_{t},d_{t})}&=\sum_{i=0}^{i_m} {P_{c_t,d_t}^{(i)} \sum_{k=0}^{b_t} T_{b_{t-1}}^{(k)} } T_{c_{t-1} }^{(c_t-c_{t-1}-i)}\nonumber\\&\quad\times S_{b_{t-1}+c_{t-1}}^{(k+b_t-b_{t-1})} {^{(1)}I_{c_{t-1}-c_{t}-i}^{(0)}}.
\end{align}
In (\ref{eq:8}) the number of group 3 users connections on cell 1 strictly increases and on cell 2 strictly decreases. All the possible terminations of group 3 users are iterated on cell 1 such that connections are generated to ensure an overall increase in group 3 users connected to cell 1. In cell 2 all the possible cases of group 3 users preemption are iterated over and the number of terminations on cell 2 are accordingly adjusted.

For $b_t=M_1,c_t+d_t \le M_2,b_t \ge b_{t-1},c_t \ge c_{t-1}$
\begin{align}
r_{b_{t-1},c_{t-1},d_{t-1}}^{(b_{t},c_{t},d_{t})}&=\sum_{i = 0}^{i_m}P_{c_t ,d_t }^{(i)} \sum_{k = 0}^{b_{t - 1}} \sum_{l=0}^{c_{t-1}}T_{b_{t - 1}}^{(k)}T_{c_{t-1}}^{(l)}\nonumber\\&\quad\times\left(S_{b_{t-1}+c_{t-1}}^{(k+l+b_t-b_{t-1}+c_t-c_{t-1}+i)}+{^{(0)}I_{i}^{(i_m)}}\right.\nonumber\\&\left.\quad\times{^{(0)}I_{c_{t}+d_{t}-i}^{(M_2)}}
\!\!\!\!\!\!\sum_{\substack{r=k+l+b_t-b_{t-1}\\+c_t-c_{t-1}+i+1}}^N\!\!\!\!\!\! S_{b_{t-1}+c_{t-1}}^{(r)} \right).
\label{eq:9}
\end{align}
The expression in (\ref{eq:9}) is similar to (\ref{eq:5}) except that the case of an increase in group 3 users connections on cell 2 is also considered. The number of generations is set to compensate for all cases of group 3 users preemption and all the possible number of terminations on both cells, to ensure the necessary increase in the number of group 3 users connections on both cells. An additional term is used to account for the group 3 users that are blocked from accessing channels when the fully-connected state is present on both cells.

Lastly, for $b_t=M_1,c_t+d_t\le M_2 ,b_t \ge b_{t-1},c_t < c_{t-1}$
\begin{align}
\label{eq:10}
r_{b_{t-1},c_{t-1},d_{t-1}}^{(b_{t},c_{t},d_{t})}&= \sum_{i=0}^{i_m} P_{c_t ,d_t }^{(i)} \left(\sum_{k=0}^{b_{t-1}}\right. \sum_{l=0}^{c_t} T_{b_{t-1}}^{(k)} T_{c_{t-1}}^{(l+c_{t-1}-c_t)}\nonumber\\&\quad\times S_{b_{t-1}+c_{t-1}}^{(k+l+b_t-b_{t-1}+c_t-c_{t-1}+i)} {^{(0)}I_i^{(0)}}\nonumber\\&\quad+\sum_{k=0}^{b_{t-1}}\sum_{l=\max\left(c_{t-1}-c_t-i,0\right)}^{c_{t-1}} T_{b_{t-1}}^{(k)} T_{c_{t-1}}^{(l)}\nonumber\\&\quad\times\left[ S_{b_{t-1}+c_{t-1}}^{(k+l+b_t-b_{t-1}+c_t-c_{t-1}+i)}{^{(1)}I_0^{(i)}}\right.\nonumber\\&\left.\left.\quad+{^{(0)}I_i^{(i_m)}}{^{(0)}I_{c_t+d_t}^{(M_2)}}\right.\right.\nonumber\\&\quad\left.\left.\times\sum_{\substack{r=k+l+b_t-b_{t-1}\\+c_t-c_{t-1}+i+1}}^N\! S_{b_{t-1}+c_{t-1}}^{(r)}\right]\right).
\end{align}
The expression in (\ref{eq:10}) is similar to (\ref{eq:7}) except that an increase in group 3 users connections on cell 1 is experienced with a decrease in the number of group 3 users connections on cell 2. Similar to (\ref{eq:9}) an additional term is present due to the fully-connected state.

\end{document}